\documentclass[12pt]{article}
\usepackage{latexsym}
\usepackage{epsfig,amssymb,euscript,slashed}
\usepackage{amsmath}
\usepackage{cite} 
\usepackage{array,calc,epsfig}
\usepackage{bbm}
\usepackage{fancybox}
\usepackage[linktocpage]{hyperref}

\def\be{\begin{equation}}
\def\ee{\end{equation}}
\def\bseq{\begin{subequations}}
\def\eseq{\end{subequations}}

\def\bea{\begin{eqnarray}}
\def\eea{\end{eqnarray}}

\def\bseq{\begin{subequations}}
\def\eseq{\end{subequations}}
\def\beq{\begin{equation}}
\def\eeq{\end{equation}}
\def\hr{\rho}

\numberwithin{equation}{section} 
\usepackage[]{graphicx}

\def\ii {{\rm i}}

\def\ket#1{|{#1}\rangle}

\def\sqr#1#2{{\vcenter{\vbox{\hrule height.#2pt
 \hbox{\vrule width.#2pt height#1pt \kern#1pt \vrule width.#2pt}\hrule
 height.#2pt}}}}

\def\slashchar#1{\setbox0=\hbox{$#1$}           
\dimen0=\wd0                                 
\setbox1=\hbox{/} \dimen1=\wd1               
\ifdim\dimen0>\dimen1                        
\rlap{\hbox to \dimen0{\hfil/\hfil}}      
#1                                        
\else                                        
\rlap{\hbox to \dimen1{\hfil$#1$\hfil}}   
/                                         
\fi}

\begin{document}
\font\cmss=cmss10 \font\cmsss=cmss10 at 7pt

\begin{flushright}{\scriptsize DFPD-15-TH-15 \\  \scriptsize QMUL-PH-15-12}
\end{flushright}
\hfill
\vspace{18pt}
\begin{center}
{\Large 
\textbf{AdS$_3$ Holography for 1/4 and 1/8 BPS geometries}
}
\end{center}

\vspace{8pt}
\begin{center}
{\textsl{ Stefano Giusto$^{\,a, b}$, Emanuele Moscato$^{\,c}$ and Rodolfo Russo$^{\,c}$}}

\vspace{1cm}

\textit{\small ${}^a$ Dipartimento di Fisica ed Astronomia ``Galileo Galilei",  Universit\`a di Padova,\\Via Marzolo 8, 35131 Padova, Italy} \\  \vspace{6pt}

\textit{\small ${}^b$ I.N.F.N. Sezione di Padova,
Via Marzolo 8, 35131 Padova, Italy}\\
\vspace{6pt}

\textit{\small ${}^c$ Centre for Research in String Theory, School of Physics and Astronomy\\
Queen Mary University of London,
Mile End Road, London, E1 4NS,
United Kingdom}\\
\vspace{6pt}

\end{center}

\vspace{12pt}

\begin{center}
\textbf{Abstract}
\end{center}

\vspace{4pt} {\small
\noindent 
Recently a new class of 1/8-BPS regular geometries in type IIB string theory was constructed in arXiv:1503.01463. In this paper we provide a precise description of the semiclassical states dual, in the AdS/CFT sense, to these geometries. In explicit examples we show that the holographic 1-point functions and the Ryu-Takayanagi's Entanglement Entropy for a single small interval match the corresponding CFT calculations performed by using the proposed dual states. We also discuss several new examples of such precision holography analysis in the 1/4-BPS sector and provide an explicit proof that the small interval derivation of the Entanglement Entropy used in arXiv:1405.6185 is fully covariant.}

\vspace{1cm}


\thispagestyle{empty}

\vfill
\vskip 5.mm
\hrule width 5.cm
\vskip 2.mm
{
\noindent  {\scriptsize e-mails:  {\tt stefano.giusto@pd.infn.it, e.moscato@qmul.ac.uk, r.russo@qmul.ac.uk} }
}

\setcounter{footnote}{0}
\setcounter{page}{0}

\newpage


\section{Introduction}\label{introduction}

Holographic dualities~\cite{Maldacena:1997re} between a $(d+1)$-dimensional gravitational theory and a $d$-dimensional Quantum Field Theory (QFT) represent a powerful tool to study quantum gravity problems. In this paper we focus on the simplest setup that was used to investigate supersymmetric black holes in 5D~\cite{Strominger:1996sh,Callan:1996dv}: the gravitational side is type IIB string theory on AdS$_3 \times S^3 \times T^4$ or AdS$_3 \times S^3 \times K3$, while the QFT dual is a $1+1$ dimensional Conformal Field Theory (CFT) with ${\cal N}=(4,4)$ supersymmetries~\cite{Maldacena:1997re}. This duality is motivated by starting with a configuration of $n_1$ D1-branes and $n_5$ D5-branes in flat space and studying a decoupling/low-energy limit. From the D-brane construction it is possible to derive the AdS radius ($R_{\rm AdS}$) in terms of the elementary string parameters and check that it is equal to the $S^3$ radius; then we can read from the geometry the central charge of the dual CFT $c=3 R_{\rm AdS}/(2 G_3) = 6 n_1 n_5$, where $G_3$ is the Newton constant of the theory reduced to AdS$_3$. The supergravity approximation is a good description of the bulk physics when curvatures are small $R_{\rm AdS}\to \infty$ which clearly corresponds to a CFT with many degrees of freedom. Moreover the gravitational description requires to work at a strongly interacting point in the moduli space of the CFT, so at a first sight it seems difficult to gain any insight on the bulk physics from the CFT side of the duality. However if we restrict the analysis to quantities that are protected by supersymmetry, then known non-renormalization theorems often imply that the results do not depend on the couplings and so can be derived by focusing on another point in the CFT moduli space where the theory is just a collection of $4 n_1 n_5$ free bosons and $4 n_1 n_5$ doublets of free chiral and anti-chiral fermions. In particular we will take advantage of the fact that a particular class of $3$-point correlators in the dual CFT is protected~\cite{Baggio:2012rr} and so can be calculated explicitly by working at the free point.

One of the aims of our analysis is to show that it is possible to use the AdS$_3$/CFT$_2$ duality to study the microstates of the Strominger-Vafa black hole, which carry D1, D5 and momentum charges. On the CFT side, the microstates that can have a dual geometric description in classical supergravity are the BPS semiclassical states with the charges of the black hole. The expectation values of the BPS operators in a semiclassical state $\ket{s_i}$ of this type give direct information on the structure of the bulk solution corresponding to $\ket{s_i}$: by using the standard AdS/CFT dictionary, each BPS operator\footnote{As explained below, for the time being we focus on operators of low dimension, even though it would be very interesting to extend the analysis further.} corresponds to a supergravity mode and so, roughly speaking, its expectation value determines a particular deviation of the microstate solution from AdS$_3 \times S^3$.

This approach was pioneered for the D1-D5 CFT in~\cite{Kanitscheider:2006zf,Kanitscheider:2007wq,Taylor:2007hs} where it was applied to $1/4$-BPS configurations, which correspond on the bulk side to the microstates of a black hole of vanishing horizon area in the supergravity limit. As we will discuss in detail, much of the technology developed in those works can be directly used also in the $1/8$-BPS case. In order to illustrate the method we will focus on the expectation values of the simplest class of BPS operators, i.e. those of (total) dimension one. The main stumbling block preventing the generalization of~\cite{Kanitscheider:2006zf,Kanitscheider:2007wq,Taylor:2007hs} to the $1/8$-BPS case has been the absence of a rich enough class of geometries with a known CFT dual. The geometries obtained by spectral flow in~\cite{Giusto:2004id,Giusto:2004ip,Ford:2006yb} have a too simple structure to highlight the general pattern, while we do not know an explicit CFT dual for the general multicentre solutions~\cite{Bena:2005va,Berglund:2005vb,Bena:2006is,Bena:2006kb}. However, recently a new class of $1/8$-BPS solutions was derived in~\cite{Bena:2015bea} with an explicit proposal for the dual semiclassical CFT states. We will focus on ``atypical'' states in this class which differ from AdS$_3 \times S^3$ already very close to the AdS boundary. Then, we have non-trivial expectation values already for BPS operators of dimension one and we can show that they match in a non-trivial way the supergravity results. This provides a strong check for the proposed dictionary between 3-charge geometries and semiclassical states.

Another interesting way to reconstruct the spacetime structure from CFT data is to study the Entanglement Entropy (EE) both on the CFT side~\cite{Calabrese:2004eu} and by using the holographic prescription of~\cite{Ryu:2006bv,Ryu:2006ef}. In particular the EE of a space interval in the CFT$_2$ probes the metric of the dual space-time deeper in the holographic direction as the interval becomes bigger. The application of this approach to the $1/4$-BPS geometries that are (small) black hole microstates was first discussed in~\cite{Giusto:2014aba} focusing on the first terms in the limit of small $l/R$ ($l$ is the size of the EE interval and $R$ is the radius of the space direction of the CFT). Again, in order to have a non-trivial match between bulk and CFT results at this order one needs to focus on ``atypical'' geometries, however this is sufficient to highlight the general issues that need to be understood in order to use the EE as a tool to characterise the microstate geometries. In general these geometries are not a metric product of (deformed) AdS$_3$ and a compact 3d space, so one needs to reformulate the extremization problem of~\cite{Ryu:2006bv,Ryu:2006ef} in terms of a codimension $2$ submanifold of a 6D geometry that is asymptotically AdS$_3 \times S^3$. A proposal on how to do this in a computationally efficient way is discussed in~\cite{Giusto:2014aba}. In this paper we show that this proposal is equivalent to the general covariant prescription of~\cite{Hubeny:2007xt} and, as an explicit application to $1/8$-BPS configurations, we test this holographic prescription for the EE in the case of the superstrata geometries derived in~\cite{Bena:2015bea}.

The paper is structured as follows. Section~2 contains a brief review of the D1-D5 CFT at the orbifold point in terms of free fields. We use this language to give a description of the BPS operators that are most relevant for our analysis, focusing in particular on those of dimension one. In Section~3 we recall the dictionary between the semiclassical CFT states and the $1/4$-BPS microstate geometries introduced in~\cite{Kanitscheider:2006zf,Kanitscheider:2007wq} and list the states we use in our concrete examples. We also extend the dictionary to the class of $1/8$-BPS microstates constructed in~\cite{Bena:2015bea}. In Section~4 the expectation values of BPS operators of dimension 1 are computed at the CFT orbifold point and on the gravity side, both for $1/4$ and for $1/8$-BPS states. In the $1/4$-BPS case, our computations generalize what has already been done in the literature in two ways: we consider states that have non-vanishing VEVs of the orbifold twist fields (for which one needs states that are made of constituents of different lengths) and highlight the meaning of the relative phases in the definition of the CFT states on the geometric side of the AdS/CFT correspondence. This more general setup makes it possible to see/check on the bulk side some fine details of the CFT twist field action recently derived in~\cite{Carson:2014ena}. In the $1/8$-BPS case, we show, in specific examples, how the match of the CFT and gravity computations crucially relies on some terms that were predicted in~\cite{Bena:2015bea} to make the bulk geometries regular. In Section~5 we study the EE of a single interval in a non-vacuum state. We show that the prescription of~\cite{Giusto:2014aba} for extracting the first state-dependent corrections in the small interval limit ($l/R\ll 1$), is equivalent to the general covariant approach of~\cite{Hubeny:2007xt}. For a generic $1/4$ or $1/8$-BPS microstate geometry we derive the EE at order ${\cal O}(l/R)^2$, both with CFT and with gravity methods, and verify the consistency between the two approaches. We comment on the relevance and the possible extension of our results for the microscopic understanding of the thermodynamic properties of black holes in Section~\ref{sec:discussion}. The appendices collect some technical results on the orbifold CFT and on the geometries dual to D1-D5 states, the computation of VEVs in $1/4$-BPS states composed of constituents of arbitrary lengths, and the proofs of some technical lemmas used in the covariant holographic EE computation. 

\section{The dual CFT at the orbifold point}
\label{sect:dCFT}

The CFT relevant for the D1-D5 microstates is a 2-dimensional QFT with ${\cal N}=(4,4)$ supersymmetry and central charge $c=6 n_1 n_5$. Here we follow the conventions of Section~7 of~\cite{Bena:2015bea} (see the references therein for more details) and visualize the CFT at the free orbifold point as a collection of $N\equiv n_1 n_5$ strings, each one with four bosons and four doublets of fermions
\begin{equation}
\Big(X^{\dot{A}A}_{(r)}(\tau,\sigma)\,, 
\psi_{(r)}^{\alpha \dot{A}}(\tau+\sigma) \,, 
\tilde\psi_{(r)}^{\dot{\alpha} \dot{A}}(\tau-\sigma)\Big) \,,
\label{Xpsipsit}
\end{equation}
where $r=1,\ldots,N$ runs over the different strings and $(\tau,\sigma)$ are the timelike and the spacelike directions in the CFT, which in our conventions will correspond to the directions $t$ and $y$ on the bulk side. Here $\alpha,\dot{\alpha} = \pm$ are spinorial indices\footnote{We sometimes use the notation $\alpha,\dot{\alpha}=1,2$ with the identifications $1\equiv +$, $2\equiv -$.} for the R-symmetry group $SU(2)_L \times SU(2)_R$ which is identified with the rotations in the non-compact space directions ($x^i$) on the bulk side, while $A,\dot{A}=1,2$ are indices for the $SU(2)_1\times SU(2)_2=SO(4)_I$ rotations acting on the tangent space in the compact manifold $T^4$. We are interested in particular in the Ramond (R) sector of the CFT, since these states correspond to geometries that can be extended to an asymptotically flat region and so correspond to the microstates of the standard black hole. The mode expansion in this sector is
\begin{equation}
  \label{psime}
  \psi_{(r)}^{\alpha \dot{A}}(\sigma^+) = \sum_{n \in Z} \psi_{n\,(r)}^{\alpha \dot{A}} {\rm e}^{-\ii n \sigma^+}~~,\qquad 
  \tilde\psi_{(r)}^{\alpha \dot{A}}(\sigma^-) = \sum_{n \in Z} \tilde\psi_{n\,(r)}^{\alpha \dot{A}} {\rm e}^{-\ii n \sigma^-}~, 
\end{equation}
where, as usual, we defined the null coordinates $\sigma^\pm = \tau \pm \sigma$. In our conventions the hermitean properties of the $\psi$'s are $\psi_{(r)\,n}^{1 \dot{1}\,\dagger} = - \psi_{(r)\,-n}^{2\dot{2}}$, $\psi_{(r)\,n}^{1 \dot{2}\,\dagger} = \psi_{(r)\,-n}^{2\dot{1}}$, and similarly for the right-moving sector. The particular R vacuum state defined by
\begin{equation}
  \label{LMvac}
  \psi_{0\,(r)}^{1 \dot{1}} \ket{++}_{(r)} = \psi_{0\,(r)}^{1 \dot{2}} \ket{++}_{(r)} = 0~,~~
  \tilde{\psi}_{0\,(r)}^{\dot{1} \dot{1}} \ket{++}_{(r)} = \tilde{\psi}_{0\,(r)}^{\dot{1} \dot{2}} \ket{++}_{(r)} =0
\end{equation}
is the building block necessary to define the CFT dual of the $1/4$-BPS microstate dubbed ``round supertube'' and in general to construct the semiclassical states we will use in our examples. 

The R-symmetry currents are obtained simply by summing over $r$ the standard currents on each string
\begin{equation}
\label{eq:Jr}
\begin{aligned}
J_{(r)}^{\alpha \beta}(\sigma^+) = & ~ \frac{1}{2} \, \psi_{(r)}^{\alpha \dot{A}}(\sigma^+) \, \epsilon_{\dot{A}\dot{B}} \,  \psi_{(r)}^{\beta \dot{B}}(\sigma^+)  \,,  
\\
\tilde J_{(r)}^{\dot{\alpha} \dot{\beta}}(\sigma^-) = & ~ \frac{1}{2}\, \tilde\psi_{(r)}^{\dot{\alpha} \dot{A}}(\sigma^-) \, \epsilon_{\dot{A}\dot{B}}  \,  \tilde\psi_{(r)}^{\dot{\beta} \dot{B}} (\sigma^-)\,,
\end{aligned}
\end{equation}
where the operators are normal-ordered with respect to the $\ket{++}_{(r)}$ vacuum and we use the convention for $\epsilon_{\dot{A}\dot{B}}$ such that $\epsilon_{\dot{1}\dot{2}}=1$. The standard $SU(2)$ generators in the R sector are $J^3_{(r)} \equiv - J^{12}_{(r)} + 1/2$ and $J^+_{(r)} \equiv J^{11}_{(r)}$, $J^-_{(r)} \equiv- J^{22}_{(r)}$. The constant term in $J^3_{(r)}$ has been fixed in such a way that the $\ket{++}_{(r)}$ state has quantum number $(1/2,1/2)$ under $(J^3_{(r)},\tilde{J}^3_{(r)})$. The currents have conformal dimension $(1,0)$ (for the $J$'s) and $(0,1)$ (for the $\tilde{J}'s)$ as usual, and we use their zero modes to define other R vacua with different spin
\begin{equation}
  \label{pmvac}
   \ket{-+}_{(r)} = J^-_{0\,(r)} \ket{++}_{(r)} \;,~~
   \ket{+-}_{(r)} = \tilde{J}^-_{0\,(r)} \ket{++}_{(r)} \;,~~
   \ket{--}_{(r)} = J^-_{0\,(r)} \tilde{J}^-_{0\,(r)} \ket{++}_{(r)} \;.
\end{equation}
There are other operators of total dimension one that will be relevant for our analysis and transform in the $(2,2)$ representation of the R-symmetry group. The first quadruplet has a simple realization in terms of free fields 
\begin{equation}
  \label{oSk}
  O^{\alpha \dot{\alpha}}_{(r)} \equiv \frac{-\ii}{\sqrt{2}} \psi^{\alpha \dot{A}}_{(r)}\, \epsilon_{\dot{A} \dot{B}} \, \tilde{\psi}^{\dot{\alpha} \dot{B}}_{(r)} =  O^{\alpha \dot{\alpha}}_{nm\,(r)} \, {\rm e}^{-\ii (n\sigma^+ + m \sigma^-)}~,
\end{equation}
which corresponds to the operator $O^{(1,1)}_{(1) 1}$ in the notation of~\cite{Kanitscheider:2007wq} and has conformal weight $(1/2,1/2)$. The action of the operator~\eqref{oSk} on the $\ket{++}$ state generates another R vacuum that plays an important role both in the examples of discussed in~\cite{Kanitscheider:2007wq} and in those of this paper
\begin{equation}
  \label{sSk}
  \ket{00}_{(r)} \equiv  \lim_{z\to 0} O^{2 \dot{2}}_{00\,(r)} \ket{++}_{(r)} = \frac{1}{\sqrt{2}} \psi^{2 \dot{A}}_{0\,(r)}\, \epsilon_{\dot{A} \dot{B}} \, \tilde{\psi}^{\dot{2} \dot{B}}_{0\,(r)} \ket{++}_{(r)}\;, 
\end{equation}
which has spin $(0,0)$ under $(J^3_{(r)},\tilde{J}^3_{(r)})$. As usual, when convenient, we pass from the cylinder coordinate for the CFT to those parametrizing a complex plane (after Wick rotation); our conventions are
\begin{equation}
  \label{plcoo}
  {\rm e}^{\ii \sigma^+} = {\rm e}^{\tau_E + \ii \sigma} = z~,~~
  {\rm e}^{\ii \sigma^-} = {\rm e}^{\tau_E - \ii \sigma} = \bar{z}\,.
\end{equation}

The other quadruplet of operators of dimension $(1/2,1/2)$ is a set of BPS twist fields $\Sigma^{\alpha \dot{\alpha}}_2$. When acting on two strings of winding one, $\Sigma^{\alpha \dot{\alpha}}_2$ joins them in a single string of winding two provided that the angular momenta~\eqref{eq:totalJ} are conserved. This means that, when going around the operator $\Sigma^{\alpha \dot{\alpha}}_2$, two copies of elementary fields in~\eqref{Xpsipsit}, say $r=1$ and $r=2$ are exchanged. Clearly, it is possible to generalize this idea and define BPS twist operators of dimension $((k-1)/2,(k-1)/2)$: they induce a cyclic permutation of $k\geq 2$ copies of elementary fields and are in a representation of spin $((k-1)/2,(k-1)/2)$ under the R-symmetry. The monodromies induced by these twists $\Sigma_{k}$ are defined by a permutation cycle specifying how the elementary fields are reshuffled when going around $\Sigma_{k}$. For each cycle of length $k$, it is convenient to diagonalize the boundary conditions in order to have $k$ independent fields. For example, if the permutation cycle involves the copies $r=1,\ldots,k$, then the monodromies of the $\Sigma_{k}$ inserted at $z=0$ are diagonalized by the combinations
\begin{equation}
   \label{eq:T2eigenvec}
   \psi_\hr^{\alpha \dot{A}} (z) = \frac{1}{\sqrt{k}}\sum_{r=1}^k {\rm e}^{-2\pi \ii \frac{r \hr}{k}}\psi_{(r)}^{\alpha \dot{A}}(z)~,~~ \mbox{with}~~~\hr=0,1,\ldots,k-1~
\end{equation}
and similarly for the other fields, which implies $\psi_\hr^{\alpha \dot{A}} ({\rm e}^{2\pi \ii} z) = {\rm e}^{2\pi \ii \hr/k}\psi_\hr^{\alpha \dot{A}} (z)$. 

The lowest weight state in the $\Sigma_k$ multiplet has spin $(-(k-1)/2,-(k-1)/2)$ and so when it acts of $k$ copies of the $\ket{++}$ vacuum produces a state $\ket{++}_k$ in the R sector of spin $(1/2,1/2)$ and winding $k$
\begin{equation}
  \label{sSi}
  \ket{++}_k \equiv \lim_{z\to 0} |z|^{k-1} \Sigma^{-\frac{k-1}{2},-\frac{k-1}{2}}_k(z,\bar{z}) \prod_{r=1}^k \ket{++}_{(r)} ~.
\end{equation}
We refer to states obtained in this way as {\em strands} of length $k$. Clearly we can change the total spin of a strand $\sum_{r=1}^k J^3_{(r)}$ by using the zero modes of $\psi_{\hr=0}^{\alpha \dot{A}}$ and similarly for the right-moving part. As in the $k=1$ case~\eqref{pmvac}, we can change the spin of the strands of length $k$ by using the currents 
\begin{equation}
  \label{pmvack}
   \ket{-+}_k = J^-_{0\,\hr=0} \ket{++}_k \;,~~
   \ket{+-}_k = \tilde{J}^-_{0\,\hr=0} \ket{++}_k \;,~~
   \ket{--}_k = J^-_{0\,\hr=0} \tilde{J}^-_{0\,\hr=0} \ket{++}_k \;
\end{equation}
or by acting with the zero mode of the operator $\sum_r O^{2 \dot{2}}_{(r)}$
\begin{equation}
  \label{sSkk}
  \ket{00}_k = \frac{-\ii}{\sqrt{2}} \psi^{2 \dot{A}}_{0\,\hr=0}\, \epsilon_{\dot{A} \dot{B}} \, \tilde{\psi}^{\dot{2} \dot{B}}_{0\,\hr=0} \, \ket{++}_k~.
\end{equation}
In Appendix~\ref{Appendix:dCFT} we collect the (standard) bosonization formula that can be used to calculate the action of $\Sigma_k$ on $\psi_\hr^{\alpha \dot{A}}$ by using free fields since we will exploit this language to calculate some of the CFT correlators in Section~\ref{CFT1p}. 

\section{Gravity-CFT map for D1-D5 states}
 \label{sec:gravityCFTmap}

The aim of this section is to precisely characterize the semiclassical states that are dual to the class of superstrata constructed in~\cite{Bena:2015bea}. We first review the CFT/geometry dictionary in the $1/4$-BPS sector by summarising the results of~\cite{Kanitscheider:2006zf,Kanitscheider:2007wq} in the language of orbifold CFT. Then we turn our attention to the $1/8$-BPS sector relevant for the superstrata.

\subsection{Gravity-CFT map in a \texorpdfstring{$1/4$-BPS}~ sector}
\label{sec_twochargesCFT}

In the previous section we introduced the concept of strands which can be used to define the states in the  D1-D5 CFT at the orbifold point. The RR ground state of each strand  is denoted by $|s\rangle_k$, where $s=(0,0), (\pm,\pm)$ runs over one of the five\footnote{We restrict here to bosonic states which are invariant under rotations of the internal space $T^4$. Hence our results trivially extend to the D1-D5 system compactified on $K_3$. If one included all the bosonic states, one would have 3 extra states for the theory on $T^4$ and 19 extra states for $K_3$. On $T^4$ there are also 8 fermionic states, while there are no fermionic states for $K_3$.} possible spin states and $k$ is the length, or winding number, of the strand. A ground state of the D1-D5 orbifold theory is obtained by taking the tensor product of $N^{(s)}_k$ copies of the strand $|s\rangle_k$, with the constraint that the total winding number be $N=n_1 n_5$.  Thus a ground state is specified by a partition $\{N^{(s)}_k\}$ of $N$:
 \begin{equation}
  \label{eq:generalgroundstate}
\psi_{\{N^{(s)}_k\}} \equiv \prod_{k,s} \,(|s\rangle_k)^{N^{(s)}_k }\,,\quad \sum_{s,k} k \,N^{(s)}_k =N\,.
\end{equation}
By convention we relate the norm of these states to the number of ways, $\mathcal{N}(\{N^{(s)}_k\})$, the strand configuration determined by the partition $\{N^{(s)}_k\}$ can be obtained starting from the state $\prod_{r=1}^N \ket{++}_{(r)} \equiv \ket{++}^N$:
\begin{equation}
\label{eq:normN}
 (\psi_{\{N^{(s)}_k\}}, \psi_{\{N'^{(s)}_k\}}) = \delta_{\{N^{(s)}_k\},\{N'^{(s)}_k\}}\, \mathcal{N}(\{N^{(s)}_k\})\,.
 \end{equation}
To compute the combinatoric factor $\mathcal{N}(\{N^{(s)}_k\})$, consider the action of the twist field $\Sigma_k^{\pm\pm}$ on $N$ copies of the CFT, to produce a strand of length $k$: there are $\frac{N!}{(N-k)!\,k}$ ways in which the twist field can act, corresponding to the possible choices of $k$ among $N$ copies, up to cyclic permutations\cite{Jevicki:1998bm}. The full state $\psi_{\{N^{(s)}_k\}}$ is obtained by acting repeatedly with twist fields, so that the total number of terms produced is
\begin{equation}
\frac{N!}{(N-k_1)!\,k_1}\frac{(N-k_1)!}{(N-k_1-k_2)!\,k_2} \ldots = \frac{N!}{\prod_{k,s} k^{N^{(s)}_k}}\,.
\end{equation}
For strands with multiplicity $N_k^{(s)}>1$, the order by which the $N_k^{(s)}$ twist operators act is immaterial, and one should hence divide by $N_k^{(s)}!$. Since each term produced by the action of twist operators has unit norm, one finds
\begin{equation}
\label{eq:normNN}
\mathcal{N}(\{N^{(s)}_k\}) = \frac{N!}{\prod_{k,s} N_k^{(s)}!\,k^{N^{(s)}_k}}\,.
\end{equation}

At the orbifold point, also the action of the operators on the CFT states contains a combinatoric part. Again this can be described in terms of permutations. The untwisted operators correspond to the identity permutation and act equally on each copy of the CFT. For instance the total angular momenta are
\begin{equation}
\label{eq:totalJ}
J^3=\sum_{r=1}^N J^3_{(r)}\,,\quad \tilde J^3 = \sum_{r=1}^N \tilde J^3_{(r)}
\end{equation}
and, by construction, the states $\psi_{\{N^{(s)}_k\}}$ are eigenstates of the zero-modes of $J^3$ and $\tilde J^3$ with eigenvalues $\sum_{k,s} s \,N^{(s)}_k$. In general the action of an operator on a D1-D5 state involves the composition of the permutation defining the operator and the permutation defining the state. Twisted operators correspond to permutations containing cycles of length $k>1$. For instance, in Section~\ref{CFT1p} we will consider the chiral primary operators with a cycle of length 2 and all others of length 1. We will still indicate them with the same symbol used in Section~\ref{sect:dCFT}, $\Sigma_2^{\pm\pm}$, understanding that one has to sum over the contributions coming from any pair of the $N$ CFT copies since the full operator contains a sum over all permutations with a single length 2 cycle.

The geometries dual to coherent superpositions of RR ground states have been constructed in \cite{Lunin:2001jy,Lunin:2002bj,Lunin:2002iz,Kanitscheider:2007wq}: they are completely specified in terms of a closed curve in $\mathbb{R}^5$, $g_A(v')$ ($A=1,\ldots,5$). The parameter along the curve, $v'$, has periodicity $L=2\pi \frac{Q_5}{R}$, where $Q_5$ is the D5 charge and $R$ is the radius of the $S^1$ on which the branes are wrapped. The equations that allow to construct the geometry given the profile $g_A(v')$, are listed in Appendix~\ref{Appendix:D1D5geometries}. The map between geometries and states can however be expressed solely in terms of the profile: the general idea is that the 5 spin states $s$ are related with the 5 components of $g_A(v')$,  the length of each strand is related with the harmonic number in the Fourier expansion of $g_A(v')$, and the magnitude of each harmonic mode specifies the number of strands of each type. More precisely, define the Fourier expansions
\begin{equation}
\label{eq:profile}
\begin{aligned}
g_1(v')+\ii\,g_2(v')  &= \sum_{n\not=0} \frac{ a^{(1)}_n}{n}\,{\rm e}^{\frac{2\pi\,\ii \,n}{L} v'} \,,~~~ 
g_3(v')+\ii\,g_3(v') &= \sum_{n\not=0} \frac{ a^{(2)}_n}{n}\,{\rm e}^{\frac{2\pi\,\ii \,n}{L} v'} \,\\
g_5(v') & = -\mathrm{Im} \Bigl[ \sum_{k=1}^\infty \frac{a^{(00)}_k}{k}\,{\rm e}^{\frac{2\pi\,\ii \,k}{L} v'}\Bigr]\,,
\end{aligned}
\end{equation}
where, for later convenience, we rename
\begin{equation}
  \label{eq:apmpm}
  a^{(1)}_{k>0} = a^{(++)}_k\;,~~  a^{(1)}_{k<0} = - a^{(--)}_{|k|}\;,~~
  a^{(2)}_{k>0} = a^{(+-)}_k\;,~~  a^{(2)}_{k<0} = - a^{(-+)}_{|k|}\;,
\end{equation}
where we highlight the contribution to the $(J^3_0,\tilde{J}^3_0)$ quantum numbers of each excitation. The Fourier coefficients $a^{(s)}_k$ are in general complex and satisfy a constraint 
\begin{equation}
\label{eq:constraint}
\sum_k \Bigl[|a^{(++)}_k|^2+|a^{(--)}_k|^2+|a^{(+-)}_k|^2+|a^{(-+)}_k|^2+\frac{1}{2}|a^{(00)}_k|^2  \Bigr] = \frac{Q_1 Q_5}{R^2}\,.
\end{equation}
The dual CFT state is more naturally expressed in terms of dimensionless coefficients $A^{(s)}_k$:
\begin{equation}
\label{eq:dimensionless}
A^{(\pm \pm)}_k \equiv R\, \sqrt{\frac{N}{Q_1 Q_5}}\,a^{(\pm \pm)}_k\,,\quad  A^{(0 0)}_k \equiv R\, \sqrt{\frac{N}{2\,Q_1 Q_5}}\,a^{(00)}_k\,,
\end{equation}
which satisfy
\begin{equation}
\label{eq:constraintbis}
\sum_{k,s} |A^{(s)}_k|^2=N \,.
\end{equation}
A given set of Fourier coefficients $\{A^{(s)}_k\}$ specifies a profile $g_A(v')$ and hence a geometry; the CFT state dual to this geometry is\cite{Skenderis:2006ah,Kanitscheider:2006zf,Kanitscheider:2007wq}
 \begin{equation}
  \label{eq:gravityCFTmap}
\psi({\{A^{(s)}_k\}}) = {\sum_{\{N^{(s)}_k\}}}^{\!\!\!\prime} (\prod_{k,s} \, A^{(s)}_k )^{N^{(s)}_k }
\psi_{\{N^{(s)}_k\}}  = {\sum_{\{N^{(s)}_k\}}}^{\!\!\!\prime} \prod_{k,s} \, (A^{(s)}_k\,|s\rangle_k)^{N^{(s)}_k }\,,
\end{equation}
where again the sum $\sum'_{\{N^{(s)}_k\}}$ is restricted to 
\begin{equation}
\label{eq:constraintN}
\sum_{s,k} k \,N^{(s)}_k =N\,.
\end{equation}  
Eq. (\ref{eq:gravityCFTmap}) gives the explicit map between gravity and CFT for states with D1, D5 charges. Notice that the states dual to geometries, $\psi({\{A^{(s)}_k\}})$, are generically superpositions of angular momentum eigenstates $\psi_{\{N^{(s)}_k\}} $. The only exception is when a single Fourier coefficient $A^{(s)}_k$ is different than zero, and hence the CFT state is composed of $N/k$ equal strands. The states whose dual geometries are well described in the classical supergravity limit are the ones in which the average numbers of strands of each type ($\overline{N}_k^{(s)}$) is very large: $\overline{N}_k^{(s)}\gg 1$. In this limit the sum over $\{N^{(s)}_k\}$ which appears in the definition of the state $\psi({\{A^{(s)}_k\}})$ is peaked over the average numbers $\overline{N}_k^{(s)}$, which are determined by the magnitudes of the Fourier coefficients $A^{(s)}_k$. To see this, consider the norm of the state $\psi({\{A^{(s)}_k\}})$:
\begin{equation}
\label{eq:norm}
|\psi({\{A^{(s)}_k\}})|^2 = {\sum_{\{N^{(s)}_k\}}}^{\!\!\!\prime} \mathcal{N}(\{N^{(s)}_k\}) \,\prod_{k,s} |A^{(s)}_k|^{2 N^{(s)}_k} \,,
\end{equation}
where we have used the orthogonality of the states $\psi_{\{N^{(s)}_k\}}$ \eqref{eq:normN}. One can now study where the sum over $\{N^{(s)}_k\}$ in (\ref{eq:norm}) is peaked in the large $N_k^{(s)}$ limit. Using the leading Stirling approximation for factorials, $\log N_k^{(s)}! \approx (N_k^{(s)} + 1/2) \log N_k^{(s)} - N_k^{(s)}$, the saddle point values $\overline{N}^{(s)}_k$ are the stationary points of the function
\begin{equation}
S(\{N^{(s)}_k\})=\sum_{k,s} N_k^{(s)} \log |A^{(s)}_k|^2-N_k^{(s)}\log N_k^{(s)} + N_k^{(s)}-N_k^{(s)}\log k\,,
\end{equation}
with the constraint $\sum_{s,k} k \,N^{(s)}_k =N$. One finds 
\begin{equation}
\label{eq:peak}
k \,\overline{N}^{(s)}_k = |A^{(s)}_k|^2\,,
\end{equation}
which is consistent with (\ref{eq:constraintbis}).

In conclusion, in the state dual to the geometry specified by the Fourier coefficients $\{A_k^{(s)}\}$, the average number of strands of type $|s\rangle_k$ is $|A^{(s)}_k|^2/k$. We will see that some properties of the geometry are sensitive not only to the average numbers $\overline{N}^{(s)}_k$, but also to the form of the state in (\ref{eq:gravityCFTmap}): in particular, the fact that the state $\psi({\{A^{(s)}_k\}})$ is a superposition of angular momentum eigenstates $\psi_{\{N^{(s)}_k\}}$ will be crucial in the following.

\subsection{Gravity-CFT map in a \texorpdfstring{$1/8$-BPS}~ sector}
\label{sec_threechargesCFT}

We saw that the profile $g_A(v')$ provides a direct link between the $1/4$-BPS geometries and the corresponding semiclassical states in the CFT. In the $1/8$-BPS sector, we do not have a complete classification of the gravitational solutions dual to states and so it is not possible to construct an exhaustive dictionary. Here we focus on the class of $1/8$-BPS geometries recently derived in~\cite{Bena:2015bea} by exploiting the linear structure of the supersymmetry equations~\cite{Bena:2011dd}. 

It is possible to construct a gravity-CFT map in this sector by relating each term in a scalar function $Z_4$ that appears in the general $1/8$-BPS ansatz (see equations (\ref{eq:generalmetric})-(\ref{eq:generalmetricbis})) to the type of strands defining the dual state. From this point of view then $Z_4$ plays the same role as the profile~\eqref{eq:profile} for the $1/4$-BPS case. We refer to Eq.~(3.20) of~\cite{Bena:2015bea} for the explicit expression of $Z_4$ in this class of solution, while here it is sufficient to say that each term in $Z_4$ is labeled by a pair $(k,m_k)$ of integer numbers satisfying $k>1$ and $0\leq m_k \leq k$ and is completely determined by a positive number $b_{k,m_k}$ and a phase $\eta_{k,m_k}$. The combination $b_{k,m_k} {\rm e}^{\ii \eta_{k, m_k}}$ plays the same role as $a_k^{(00)}$ in~\eqref{eq:profile}. 

In analogy with the discussion of the $1/4$-BPS case, we define the following eigenstates of total angular momenta~\eqref{eq:totalJ}
 \begin{equation}
  \label{eq:o3cstate}
\psi_{\{N^{(s)}_{k,m_k}\}} \equiv \prod_{s=1}^4 \prod_{k}  \left(|s\rangle_k\right)^{N^{(s)}_k }  \prod_{k,m_k}  \left(\frac{(J^+_{-1})^{m_k}_k}{m_k!}|00\rangle_k\right)^{N^{(00)}_{k,m_k} }
\end{equation}
where $s=1, \ldots,4$ corresponds to the strands $\ket{\pm\pm}_k$, $(J^+_{n})_k$ is the current acting on a strand of length $k$ and, as before, the sum is constrained by~\eqref{eq:constraintN}. The states represent a generalization of the $1/4$-BPS building block in~\eqref{eq:generalgroundstate} because we now allow for the presence of RR ground states $\ket{00}_k$ excited with $m_k \leq k$ insertions of $(J^+_{-1})_k$ (it can be checked by using the free field representation of Section~\ref{sect:dCFT} that $m_k$ cannot be greater than $k$ otherwise the state vanishes). Then the $(0,0)$ strands in~\eqref{eq:o3cstate} have eigenvalue $m_k$ for both $(L_0)_k$ and $(J^3_0)_k$. The normalization $\mathcal{N}(\{N^{(s)}_{k,m_k}\})$ of these states is related to the combinatoric properties of the permutation $\{N^{(s)}_{k,m_k}\}$ but contains also an extra factor derived from the contractions of the $(J^+_{-1})_k$ insertions
\begin{equation}
\label{eq:normNN3c}
\mathcal{N}(\{N^{(s)}_{k,m_k}\}) = 
\left( \frac{N!}{\prod_{s=1}^4 \prod_{k} N_k^{(s)}!\,k^{N^{(s)}_k}}\right) \left( \frac{1}{\prod_{k,m_k} N_{k,m_k}^{(00)}!\,k^{N^{(00)}_{k,m_k}}}\right) \prod_{k.m_k}
\binom{k}{m_k}^{N_{k,m_k}^{(00)}}\,.
\end{equation}

Then we can define the states $\psi({\{A^{(s)}_k,B_{k,m_k}\}})$ as follows
 \begin{equation}
  \label{eq:gravityCFTmap3c}
\psi({\{A^{(s)}_k,B_{k,m_k}\}}) = {\sum_{\{N^{(s)}_{k,m_k}\}}}^{\!\!\!\prime} \left[\prod_{s=1}^4 \prod_{k}  (A^{(s)}_k\,|s\rangle_k)^{N^{(s)}_k } \prod_{k,m_k}  \left(B_{k,m_k} \frac{(J^+_{-1})^{m_k}_k}{m_k!}|00\rangle_k\right)^{N^{(00)}_{k,m_k} }\right]\,,
\end{equation}
with norm
\begin{equation}
\label{eq:norm3c}
|\psi({\{A^{(s)}_k,B_{k,m_k}\}})|^2 = {\sum_{\{N^{(s)}_{k,m_k}\}}}^{\!\!\!\prime} \mathcal{N}(\{N^{(s)}_{k,m_k}\}) \,\left(\prod_{s=1}^4 \prod_{k} |A^{(s)}_k|^{2 N^{(s)}_k}\right) \left( \prod_{k,m_k} |B_{k,m_k}|^{2N_{k,m_k}^{(00)}} \right) \,.
\end{equation}
The numbers of strands $\overline{N}^{(s)}_{k,m_k}$ on which the sum in (\ref{eq:norm3c}) is peaked are the stationary points of the function
\begin{align}
S(\{N^{(s)}_{k, m_k}\})=&\sum_{s=1}^4 \sum_{k}\left[ N_k^{(s)}\log |A^{(s)}_k|^2-N_k^{(s)}\log N_k^{(s)} + N_k^{(s)}-N_k^{(s)}\log k\right] +\nonumber\\
&+\sum_{k,m_k} \Bigl[ N_{k,m_k}^{(00)} \log |B_{k,m_k}|^2  - N_{k,m_k}^{(00)} \log N_{k,m_k}^{(00)} + N_{k,m_k}^{(00)} - N_{k,m_k}^{(00)} \log k+ \nonumber\\
&  + N_{k,m_k}^{(00)} \log \binom{k}{m_k} \Bigr] \,,
\end{align}
again with the constraint $\sum_{s,k} k \,N^{(s)}_k + \sum_{k,m_k} k N_{k,m_k}^{(00)} =N$. One finds 
\begin{equation}
\label{eq:peak3c}
k \,\overline{N}^{(s)}_k = |A^{(s)}_k|^2,\qquad k \, \overline{N}_{k,m_k}^{(00)} = \binom{k}{m_k} |B_{k,m_k}|^2 \,.
\end{equation}
We can relate the coefficients $A^{(s)}_k$ with $s=(\pm,\pm)$ to the supergravity parameters $a^{(s)}_k$ by using~\eqref{eq:dimensionless}, while for $s=(00)$ we have
\begin{equation}
\label{eq:dimensionless3c}
B_{k,m_k} \equiv R\, \sqrt{\frac{N}{2\,Q_1 Q_5}}\,\binom{k}{m_k}^{-1}\, b_{k, m_k} {\rm e}^{\ii \eta_{k, m_k}}\,.
\end{equation}
Note that the gravity parameters $a\equiv a^{(++)}_1$ and $b_{k,m_k}$ satisfy the constraints (6.10) in \cite{Bena:2015bea}, which generalizes the constraint (\ref{eq:constraint}) valid for two-charge geometries. When translated in terms of the CFT parameters $A^{(s)}_k$ and $B_{k,m_k}$, using the above dictionary, the constraint becomes 
\begin{equation}
\label{eq:constraintbis3charge}
\sum_{s=1}^4 \sum_k |A^{(s)}_k|^2 + \sum_k  |B_{k,m_k}|^2=N \,,
\end{equation}
which generalizes (\ref{eq:constraintbis}). 

\section{CFT 1-point functions and holography}
\label{CFT1p}

Holography allows to extract the 1-point functions of chiral primary operators in 1/4 and 1/8 BPS states from the asymptotic expansion of the dual geometries. As these 1-point functions are protected, they should match the VEVs computed at the free orbifold point of the CFT. We concentrate in this section on chiral primaries of dimension 1 and work out a series of examples that confirm the gravity-CFT map defined in the previous section. 

We start by recalling the connection between the geometry and the VEVs of CFT operators for a general D1-D5-P microstate \cite{Kanitscheider:2006zf,Kanitscheider:2007wq}. The 6D Einstein frame metric for such a microstate can be written in the form\cite{Giusto:2013rxa}
\begin{equation}
\label{eq:generalmetric}
ds^2_6 = - \frac{2}{\sqrt{\mathcal{P}}} (dv +\beta) \left( du + \omega + \frac{\mathcal{F}}{2}(dv + \beta) \right) + \sqrt{\mathcal{P}}\,ds_4^2,
\end{equation}
with
\begin{equation}
\label{eq:generalmetricbis}
\mathcal{P} = Z_1 Z_2 - Z_4^2\,.
\end{equation}
We have introduced here the light-cone coordinates
\begin{eqnarray}
u = \frac{t-y}{\sqrt{2}}\quad,\quad v = \frac{t+y}{\sqrt{2}}\,,
\end{eqnarray}
constructed from the time coordinate $t$ and the $S^1$ coordinate $y$. The remaining four spatial coordinates $x^1, \ldots, x^4$ form a Euclidean space with metric $ds^2_4$; though this metric is non-flat, generically, it reduces to flat $\mathbb{R}^4$ asymptotically. The remaining ingredients encoding the 6D metric are the four scalars $Z_1$, $Z_2$,  $Z_4$ and $\mathcal{F}$, and the 1-forms on $\mathbb{R}^4$ $\beta$ and $\omega$. For general 3-charge geometries all these scalars and 1-forms depend on $v$ and on $x^i$. For 2-charge geometries the $v$ dependence disappears and $\mathcal{F}=0$.

At leading order in the large distance expansion the metric (\ref{eq:generalmetric}) reduces to AdS$_3\times S^3$. To extract the VEVs of operators of dimension 1, it is enough to keep the first non-trivial corrections around AdS$_3\times S^3$, which have the form
\begin{subequations}
\label{eq:asymptotics}
\begin{align}
Z_1 &= \frac{Q_1}{r^2} \left( 1 + \frac{f^1_{1i}}{r} Y^i_1 + O(r^{-2})\right)\,,\quad
Z_2 = \frac{Q_5}{r^2} \left( 1 + \frac{f^5_{1i}}{r} Y^i_1 + O(r^{-2})\right)\,,\label{eq:asymptotics1}\\
Z_4 &= \frac{\sqrt{Q_1 Q_5}}{r^3} \mathcal{A}_{1i} Y^i_1 + O(r^{-4})\,,\quad \mathcal{F} = -\frac{2\,Q_p}{r^2} + O(r^{-3})\,,\quad ds^2_4 = dx^i dx^i + O(r^{-4})\,,\\
 \beta&= -\frac{\sqrt{2\,Q_1 Q_5}}{r^2} a_{\alpha-} Y^{\alpha-}_1 + O(r^{-3})\,,\quad \omega = -\frac{\sqrt{2\,Q_1 Q_5}}{r^2} a_{\alpha+} Y^{\alpha+}_1 + O(r^{-3}) \,.
\end{align}
\end{subequations}

It is always possible to pick coordinates in such a way that 
\begin{equation}
\label{eq:gauge}
f^1_{1i}+f^5_{1i}=0\,,
\end{equation}
and we will always assume this gauge choice in the following. We have denoted by $Y^i_1$ the $l=1$ scalar spherical harmonics on $\mathbb{R}^4$, and by $Y_1^{\alpha\pm}$ the $l=1$ vector spherical harmonics; their expressions are\begin{align}
Y^i_1 &= 2\frac{x^i}{r}\,,\quad Y_1^{\alpha+} = \frac{\eta^\alpha_{ij} dx^i x^j}{r^2}\,,\quad Y_1^{\alpha-} = \frac{\bar{\eta}^\alpha_{ij} dx^i x^j}{r^2}\,,
\end{align}
where $\eta^\alpha_{ij} = \delta_{\alpha i} \delta_{4j} - \delta_{\alpha j} \delta_{4i} + \epsilon_{\alpha ij4}$ and $\bar{\eta}^\alpha_{ij} = \delta_{\alpha i} \delta_{4j} - \delta_{\alpha j} \delta_{4i} - \epsilon_{\alpha ij4}$ (with $\alpha = 1,2,3$) are the 't Hooft symbols. $Q_1$, $Q_5$ and $Q_p$ are the D1, D5 and P charges, which are quantized in terms of positive integers $n_1$, $n_5$, $n_p$ as
\begin{equation}
\label{chargequantization}
Q_1 = \frac{(2\pi)^4 \,n_1\,g_s\,\alpha'^3}{V_4}\,,\quad Q_5 = n_5\,g_s\,\alpha'\,,\quad Q_p = \frac{(2\pi)^4\, n_p\,g_s^2\,\alpha'^4}{R^2\, V_4}\,,
\end{equation}
with $g_s$ the string coupling, $R$ the radius of $S^1$ and $V_4$ the volume of $T^4$. The coefficients $f^1_{1i}$, $\mathcal{A}_{1i}$, $a_{\alpha\pm}$ are constants for 
2-charge geometries but might depend on the light-cone coordinate $v$ for 3-charge states. They capture the VEVs of the chiral primaries of conformal dimension 1.

These chiral primaries comprise the $SU(2)_L \times SU(2)_R$ currents $J^3$ and $\tilde{J}^3$ (which have dimensions $(1,0)$ and $(0,1)$), and the operators of dimension $(1/2,1/2)$,  $\Sigma_2^{\alpha\dot\alpha}$ and $O^{\alpha\dot\alpha}$, introduced in Section~\ref{sect:dCFT}; it is understood that these operators contain a sum over all copies of the CFT, as in \eqref{eq:totalJ}. The same operators where introduced in \cite{Kanitscheider:2007wq}, where they were denoted by $O^{(0,0)}_{(2)i}$ and $O^{(1,1)}_{(1)1i}$; we give here an explicit representation of the operators at the free orbifold point of the CFT. The precise relation between our operators and the operators of \cite{Kanitscheider:2007wq} is\footnote{The minus sign in the second equations in (\ref{eq:Sigma2}) and (\ref{eq:OO}) is imposed by consistency with the $SU(2)$ algebra.}
\begin{subequations}
\label{eq:sigma2++--}
\begin{align}
\Sigma_2^{++} &= O^{(0,0)}_{(2)1} +\ii O^{(0,0)}_{(2)2}\,,\quad \Sigma_2^{--} = (\Sigma_2^{++})^\dagger= O^{(0,0)}_{(2)1} -\ii O^{(0,0)}_{(2)2}\,,\\
\Sigma_2^{+-} &= O^{(0,0)}_{(2)3} +\ii O^{(0,0)}_{(2)4}\,,\quad \Sigma_2^{-+} =-(\Sigma_2^{+-})^\dagger=-\left( O^{(0,0)}_{(2)3} -\ii O^{(0,0)}_{(2)4}\right)\,,\label{eq:Sigma2}
\end{align}
\end{subequations}
and similarly
\begin{subequations}
\label{eq:O++--}
\begin{align}
O^{++} &= O^{(1,1)}_{(1)11} + \ii O^{(1,1)}_{(1)12}\,,\quad O^{--} = (O^{++})^\dagger=O^{(1,1)}_{(1)11} - \ii O^{(1,1)}_{(1)12}\,,\\
O^{+-} &= O^{(1,1)}_{(1)13} + \ii O^{(1,1)}_{(1)14}\,,\quad O^{-+} =-(O^{+-})^\dagger=-\left( O^{(1,1)}_{(1)13} - \ii O^{(1,1)}_{(1)14}\right)\,.\label{eq:OO}
\end{align}
\end{subequations}

The relation between the 1-point functions of these operators in a state $|s\rangle$ and the dual geometry was worked out in \cite{Kanitscheider:2006zf,Kanitscheider:2007wq}, and it is given by
\begin{subequations}
\label{gravity_cft_vev}
\begin{align}
\langle s| J^\alpha |s\rangle &= c_J\, a_{\alpha+}\quad ,\quad \langle s| \tilde{J}^\alpha |s\rangle = c_{\tilde{J}}\, a_{\alpha-}
\quad ,\quad \langle s| L_0 - \tilde{L}_0 |s\rangle = n_p\,,
\label{gravity_cft_vev_j}\\
& \langle s| O^{(0,0)}_{(2)i} |s\rangle = c_{O^{(0,0)}}\, f^1_{1i}\quad ,\quad \langle s| O^{(1,1)}_{(1)1i} |s\rangle  = c_{O^{(1,1)}}\, \mathcal{A}_{1i}\,.\label{gravity_cft_vev_twist}
\end{align}
\end{subequations}

The coefficients $c_j$, $c_{\tilde J}$, $c_{O^{(0,0)}}$, $c_{O^{(1,1)}}$ are constants {\it independent of the state}; their value is difficult to determine a priori, and hence we will fix them by comparison with some particular simple state. We will see that consistency between the CFT and the holographic computations of the entanglement entropy in the D1-D5 microstates provides a non-trivial check on the values of these coefficients. In \cite{Giusto:2014aba} this consistency relation was used to fix some of these coefficients:
\begin{align}
\label{eq:cvalues}
c_J = - c_{\tilde{J}} = \frac{NR}{\sqrt{Q_1 Q_5}},\qquad  & c_{O^{(1,1)}} = \frac{\sqrt{2}\,NR}{\sqrt{Q_1 Q_5}}\,;
\end{align}
as expected, they only depend on the asymptotic moduli.

All microstates considered in previous works had vanishing VEVs of the twist operators $\Sigma_2^{\alpha\dot{\alpha}}$, and hence the coefficient $c_{O^{(0,0)}}$ was left undetermined. One of the purposes of the next section is to fill this gap, by considering a microstate where the VEV of $\Sigma_2^{\alpha\dot{\alpha}}$ is non-trivial.

\subsection{Switching on the twist fields' VEVs}
\label{Section:twist}
In this section we analyze the simplest D1-D5 microstate in which the VEV of the twist field $\Sigma_2^{\alpha\dot{\alpha}}$ is non-vanishing. Since the twist field can join two strands of winding one into a strand of winding two (or split a doubly wound strand into two singly wound strands), a state which contains both strands of winding one and two has a non-trivial $\Sigma_2^{\alpha\dot{\alpha}}$ VEV. A more general situation, in which the twist field joins strands of winding $k_1$ and $k_2$ into a strand of winding $k_1+k_2$ will be considered in Appendix~\ref{Appendix:twist}.

The building blocks of the state we consider here are the strands $\ket{++}_{k=1}$ and $\ket{++}_{k=2}$, where $\ket{++}_{k}$ is defined in \eqref{sSi}. As we explained in Section~\ref{sec:gravityCFTmap}, to have a state which is well described by a classical geometry one needs to take a linear superposition of states of the form (\ref{eq:gravityCFTmap}), where now only the coefficients $A_1^{(++)}$ and $A_2^{(++)}$ are non-vanishing; $N_1^{(++)}$ and $N_2^{(++)}$ denote the numbers of strands of type $\ket{++}_{k=1}$ and $\ket{++}_{k=2}$. To lighten the notation, in this section we rename $A_1^{(++)}\equiv A_1$, $A_2^{(++)}\equiv A_2$ and $N_2^{(++)}\equiv p$. Then the constraint (\ref{eq:constraintN}) implies $N_1^{(++)} = N-2p$. The state we consider is then
\begin{equation}
\label{eq:stateAB}
\psi (A_1,A_2) = \sum_{p=1}^{N/2} \bigl( A_1 \ket{++}_1\bigr)^{N-2 p}  \bigl( A_2 \ket{++}_2\bigr)^{p}\,,
\end{equation}
where
\begin{equation}
|A_1|^2 + |A_2|^2 = N\,,
\end{equation}
as a consequence of \eqref{eq:constraintbis}. We know from (\ref{eq:peak}) that the sum in (\ref{eq:stateAB}) is peaked over 
\begin{equation}
\label{eq:peakp}
\bar{p} \equiv \overline{N}_2^{(++)}= \frac{|A_2|^2}{2}\quad\Rightarrow \quad \overline{N}_1^{(++)} = N- 2 {\bar p}=|A_1|^2\,.
\end{equation}
Note that the state $\psi(A_1,A_2)$ is not normalized, but its norm is
\begin{equation}
\label{eq:normAB}
| \psi(A_1, A_2)|^2 = \sum_{p=1}^{N/2} \mathcal{N}(p)\, |A_1|^{2(N-2p)} |A_2|^{2p} \quad \mathrm{with}\quad \mathcal{N}(p) = \frac{N!}{(N-2p)!\,p!\,2^{p}}\,,
\end{equation}
where we have used \eqref{eq:norm}.

By conservation of the angular momenta $J^3$ and $\tilde{J}^3$ it is easi to determin which of the operators $\Sigma_2^{\alpha\dot{\alpha}}$ acquire a VEV in the above state. When $\Sigma_2^{\alpha\dot{\alpha}}$ acts on two strands of type $\ket{++}_{k=1}$, it produces a state with winding two and  angular momenta $\left( 1 + \alpha/2, 1 + \dot{\alpha}/2 \right)$, with $\alpha, \dot{\alpha} = \pm 1$; for the VEV of the twist field to be non-zero, this latter state has to overlap with the state $\ket{++}_{k=2}$, whose spin is $\left( 1/2 , 1/2 \right)$. One thus needs $\alpha=\dot\alpha = -1$, which means that $\Sigma_2^{--}$ acquires VEV in the state (\ref{eq:stateAB}). Since $\Sigma_2^{++}= (\Sigma_2^{--})^\dagger$, the VEV of $\Sigma_2^{++}$ must also be non-zero: this VEV originates from the process in which $\Sigma_2^{++}$ acts on the doubly wound strand $\ket{++}_{k=2}$ to produce two copies of the singly wound strand $\ket{++}_{k=1}$. 

Consider first the VEV of $\Sigma_2^{--}$: the relevant contribution comes from the process in which the twist field lowers by two the number of length one strands and increases by one the number of length two strands, which is represented by
\begin{equation}
\label{eq:Sigma2action}
\Sigma_2^{--} \left[ \bigl( \ket{++}_1 \bigr)^{N-2p} \bigl( \ket{++}_2 \bigr)^{p} \right] = (p+1) \left[ \bigl( \ket{++}_1 \bigr)^{N-2(p+1)} \bigl( \ket{++}_2 \bigr)^{p+1} \right]\,.
\end{equation}
The combinatorial factor $p+1$ can be understood as follows. The twist field $\Sigma_2^{--}$ can act on any one of the $\binom{N-2p}{2}$ copies of length one strands in the state
$\left[ \bigl( \ket{++}_1 \bigr)^{N-2p} \bigl( \ket{++}_2 \bigr)^{p} \right]$, which is made of $\mathcal{N}(p)$ terms; the total number of terms on the l.h.s. and the r.h.s. of (\ref{eq:Sigma2action}) matches if
\begin{equation}
\binom{N-2p}{2} \,\mathcal{N}(p) = (p+1)\,\mathcal{N}(p+1)\,, 
\end{equation}
which is verified using the expression for $\mathcal{N}(p)$ in (\ref{eq:normAB}).

From the basic action (\ref{eq:Sigma2action}), one therefore has
\begin{equation}
\Sigma_2^{--} \psi ( A_1,A_2 ) = \sum_{p=1}^{N/2} A_1^{N-2p} A_2^p\, (p+1) \bigl( \ket{++}_1\bigr)^{N-2(p+1)}  \bigl( \ket{++}_2\bigr)^{p+1}
\end{equation}

The VEV of $\Sigma_2^{--}$ over $\psi(A_1, A_2)$ is then computed as
\begin{align}
\label{eq:VEVSigma2}
\langle \Sigma_2^{--} \rangle&\equiv  | \psi(A_1, A_2)|^{-2}\,\langle \psi(A_1,A_2)| \Sigma_2^{--}\ket{\psi(A_1,A_2)} \nonumber\\
&= \frac{A_1^2}{A_2}\,| \psi(A_1, A_2)|^{-2} \sum_{p=1}^{N/2} \left(|A_1|^2\right)^{N-2p} \left(|A_2|^2\right)^p p \,\mathcal{N}(p) = \frac{A_1^2}{A_2}\ \bar{p} = \frac{A_1^2 \bar{A_2}}{2}\,,
\end{align}
where, in the last step, we have used (\ref{eq:peakp}).

For consistency, we should also verify that the VEV of $\Sigma_2^{++}$ is the complex conjugate of the VEV in (\ref{eq:VEVSigma2}). The relevant action of $\Sigma_2^{++}$ is given by
\begin{equation}\label{eq:Sigma2daggeraction}
\Sigma_2^{++} \left[ \bigl( \ket{++}_1 \bigr)^{N-2p} \bigl( \ket{++}_2 \bigr)^{p} \right] = \frac{(N-2p+1)(N-2p+2)}{2} \left[ \bigl( \ket{++}_1 \bigr)^{N-2p+2} \bigl( \ket{++}_2 \bigr)^{p-1} \right],
\end{equation}
where the combinatorial factor follows from the identity
\begin{equation}\label{eq:identitySigma2dagger}
p\,\mathcal{N}(p) = \frac{(N-2p+1)(N-2p+2)}{2} \,\mathcal{N}(p-1)\,,
\end{equation}
which can be derived by following steps similar to those explained after~(\ref{eq:Sigma2action}); note that the factor $p$ on the l.h.s. of the above equation comes from the $p$ possible ways in which  $\Sigma_2^{++}$ can act on the $p$ strands of type $\ket{++}_2 $. It follows by comparison of (\ref{eq:Sigma2action}) and (\ref{eq:Sigma2daggeraction}), and by the identity (\ref{eq:identitySigma2dagger}),  that
\begin{equation}
(\psi(p+1), \Sigma_2^{--}\, \psi(p) ) =(\Sigma_2^{++} \,\psi(p+1), \psi(p))\,,
\end{equation}
where for brevity we have denoted
\begin{equation}
\psi(p) \equiv   \left[ \bigl( \ket{++}_1 \bigr)^{N-2p} \bigl( \ket{++}_2 \bigr)^{p} \right] \,.
\end{equation}
This proves that indeed $\Sigma_2^{++} = (\Sigma_2^{--})^\dagger$ and it implies that 
\begin{align}
\langle \Sigma_2^{++}\rangle  =\langle \Sigma_2^{--}\rangle^* = \frac{\bar{A_1}^2 A_2}{2}.
\end{align}

The only other operators of dimension one that have a non-vanishing VEV in the state $\psi(A_1,A_2)$ are the currents $J^3$, $\tilde J^3$. These VEVs can be straightforwardly computed, as they are only sensitive to the average numbers of strands of length one and two, which both carry spin $(1/2,1/2)$. Using (\ref{eq:peakp}) one then finds
\begin{equation}
\label{eq:CFTJs}
\langle J^3 \rangle = \langle \tilde J^3 \rangle = \frac{1}{2} \left(\overline{N}_1^{(++)} + \overline{N}_2^{(++)}\right) = \frac{1}{2}\left( |A_1|^2+\frac{|A_2|^2}{2}\right)\,.
\end{equation}

We now compare the 1-point functions computed in the CFT with the ones extracted from the dual geometry. This is the geometry associated with a profile whose only two excited modes are $a_1^{(++)}$ and $a_2^{(++)}$, in the notation of (\ref{eq:profile}). For notational simplicity we abbreviate $a_1^{(++)}\equiv a_1$ and $a_2^{(++)}\equiv a_2$. The relation between $a_1$, $a_2$ and $A_1$, $A_2$ is given in \eqref{eq:dimensionless}: 
\begin{align}
a_i = \frac{A_i}{R} \sqrt{\frac{Q_1 Q_5}{N}}\quad (i=1,2)\,.
\end{align}

The parameters which encode the asymptotic behavior of the geometry, defined in general in (\ref{eq:asymptotics}), take the following values for our microstate (see Appendix~\ref{Appendix:D1D5geometries}):
\begin{align}
\label{eq:gravitytwist}
f^1_{11} - \ii f^1_{12} &=  \frac{R^2}{2\,Q_1 Q_5} a_1^2 {\bar a}_2\,,\quad \mathcal{A}_{1i} = 0\,,\quad a_{3+} = -a_{3-} = \frac{R}{2 \,\sqrt{Q_1 Q_5}} \left( |a_1|^2 + \frac{|a_2|^2}{2}\right)\,.
\end{align}

Using the dictionary in (\ref{gravity_cft_vev_j}), with the $c_J$ and $c_{\tilde J}$ of (\ref{eq:cvalues}), one readily verifies that the VEVs of $J_3$ and $\tilde J_3$ computed in (\ref{eq:CFTJs}) agree with their holographically derived values. For the VEV of $\Sigma_2^{--}$, the first of (\ref{gravity_cft_vev_twist}), together with (\ref{eq:sigma2++--}), gives
\begin{equation}
\langle \Sigma_2^{--} \rangle = c_{O^{(0,0)}}\, (f^1_{11} - \ii f^1_{12}) \,.
\end{equation}
Comparison of the CFT (\ref{eq:VEVSigma2}) and gravity (\ref{eq:gravitytwist}) results fixes the value of the unknown coefficient $c_{O^{(0,0)}}$:
\begin{equation}
\label{twist_constant}
c_{O^{(0,0)}} = \frac{N^{3/2}R}{\sqrt{Q_1 Q_5}}.
\end{equation}
The fact that $c_{O^{(0,0)}}$ is independent of $a_1$, $a_2$ represents already a non-trivial check; we will see that the precise numerical value of $c_{O^{(0,0)}}$ is checked also by the computation of the entanglement entropy in the state $\psi(A_1,A_2)$. 

\subsection{3 charges and two kinds of strands}\label{subsection:3c_2st}
We now extend the holographic computation of 1-point functions of dimension 1 chiral primaries to the class of three-charge microstates introduced in Section~\ref{sec_threechargesCFT}. Consider first a simple D1-D5-P state containing only two types of strands: strands of type $\ket{++}_{k=1}$ in their ground state and strands of type $\ket{00}_{k=1}$ acted upon by the current  $J_{-1}^+$, which carries momentum. The geometry dual to this state was first constructed in \cite{Giusto:2013rxa,Giusto:2013bda}. The CFT state has the form (\ref{eq:o3cstate}) where the non-vanishing coefficients are $A_1^{(++)}$ and $B_{1,1}$; renaming $A_1^{(++)}\equiv A$, $B_{1,1} \equiv B$ and $N_1^{(++)}\equiv p$, and using the constraint \eqref{eq:generalgroundstate}, we get
\begin{equation}
\psi ( A, B) \equiv \sum_{p=0}^N \left( A \ket{++}_1\right)^{p} \left( B\ J_{-1}^+ \ket{00}_1\right)^{N-p}.
\end{equation}
The constraint \eqref{eq:constraintbis3charge} now reads
\begin{equation}
\label{constr_chi}
|A|^2 + |B|^2 = N.
\end{equation}
and from \eqref{eq:peak3c} we have
\begin{equation}
\overline{N}_1^{(++)} = \bar{p} = |A|^2,\qquad \overline{N}_1^{(00)} = N-\bar{p} = |B|^2.
\end{equation}
These relations immediately give the VEVs of the angular momentum operators:
\begin{equation}
\langle J^3 \rangle =  \frac{\overline{N}_1^{(++)}}{2} + \overline{N}_1^{(00)} = \frac{|A|^2}{2} + |B|^2,\qquad  \langle \tilde{J}^3 \rangle = \frac{\overline{N}_1^{(++)}}{2} =  \frac{|A|^2}{2}\,,
\end{equation}
since the strands $\ket{++}_{k=1}$ and $J^+_{-1}\ket{00}_{k=1}$ carry angular momenta $(1/2,1/2)$ and $(1,0)$ respectively. We can also read off the average value of momentum:
 \begin{equation}
\langle \tilde{T} \rangle = 0\;,~~~
\langle T \rangle = (N-\bar p) = |B|^2 ~~\Rightarrow~~
 n_p = \langle L_0 - \tilde{L}_0  \rangle = |B|^2\,,
\end{equation}
since every strand $J_{-1}^+ \ket{00}_1$ carries $1$ unit of momentum.

Consider now the operator $O^{\alpha\dot{\alpha}}$. As one sees from (\ref{sSk}) the operator $O^{2\dot{2}}$ transform the strand $\ket{++}_1$ into $\ket{00}_1$;  in our state, the $\ket{00}_1$ strand is acted upon by $J^+_{-1}$, and thus, to determine the action of $O^{\alpha\dot{\alpha}}$ on the state $\psi(A,B)$ we need to know the commutation properties of $O^{\alpha\dot{\alpha}}$ with the $SU(2)$ current algebra. As the index $\alpha$ transforms in the fundamental representation of $SU(2)$ (which we represent by the matrices $\tau^i = \sigma^i/2$), one has the following nontrivial commutator\footnote{A similar relation holds for the modes of $\tilde{J}^3$, with the difference that the dotted index gets rotated.}
\begin{align}
\label{eq:commutatorJ}
\left[ \left(J^i_n\right)^{\alpha\beta}, O^{\beta\dot{\alpha}}(v, u)\right] = \frac{1}{2} {\rm e}^{\ii n\frac{\sqrt{2}v}{R}} \left(\sigma^i\right)^{\alpha\beta} O^{\beta\dot{\alpha}}(v,u),
\end{align}
where the $v$-dependent factor comes from the fact that we are considering the $n$-th mode of current $J^i(v,u)$. Hence if we use (\ref{sSk}), the commutator (\ref{eq:commutatorJ}) and the fact that positive modes of the currents annihilate the vacuum strands ($(J^i_n)_k \ket{s}_k=0$ for $n>0$), we obtain the following VEVs for individual strands
 \begin{align}
\phantom 1_1 \langle 00| J^-_{+1} \,O^{1\dot{2}}(v,u) \ket{++}_1 = {\rm e}^{\ii\frac{\sqrt{2}v}{R}}\,,\quad \phantom 1_1 \langle ++| O^{2\dot{1}}(v,u) \,J^+_{-1}  \ket{00}_1 = -{\rm e}^{-\ii\frac{\sqrt{2}v}{R}},
\end{align}
which are consistent with the hermiticity property $O^{2\dot{1}} = -\left( O^{1\dot{2}}\right)^\dagger$.  Note that it is important that the operators $O^{1\dot{2}}$ and  $O^{2\dot{1}}$ are inserted at a generic worldsheet point  $(v,u)$ and that, due to the presence of the current $J^-_{+1}$, $J^+_{-1}$, the non-zero-mode part of the operators contributes to the correlator; if only zero-modes had contributed, $O^{2\dot{1}}$ would have annihilated the state $\ket{++}_1$ because of (\ref{LMvac}).

The action of $O^{2\dot{1}}$ on angular momentum eigenstates is obtained by combining the above result with the appropriate combinatorial factor\footnote{$O^{2\dot{1}}$ can act on the $N-p$ strands of type $J_{-1}^+ \ket{00}_1$, producing $(N-p)\binom{N}{p} =(p+1) \binom{N}{p+1}$ terms, which matches the number of terms on the r.h.s. of (\ref{eq:O21action}). }
\begin{equation}
\label{eq:O21action}
O^{2\dot{1}}(v,u) \biggl[ \left(  \ket{++}\right)^{p}_1 \left( J_{-1}^+ \ket{00}_1\right)^{N-p} \biggr]= -{\rm e}^{-\ii\frac{\sqrt{2}v}{R}}\,(p+1)\biggl[ \left(  \ket{++}\right)^{p+1} \left( J_{-1}^+ \ket{00}\right)^{N-p-1}\biggr]\,.
\end{equation}
The VEV of $O^{2\dot{1}}$ on the state $\psi(A,B)$ is then
\begin{align}
\langle O^{2\dot{1}}(v,u) \rangle &= -{\rm e}^{-\ii\frac{\sqrt{2}v}{R}} \frac{B}{A} \,\bar{p}= -{\rm e}^{-\ii\frac{\sqrt{2}v}{R}} \bar{A} B.
\end{align}
Because $O^{2\dot{1}} = -\left( O^{1\dot{2}}\right)^\dagger$, the VEV of $O^{1\dot{2}}$ is
\begin{equation}
\langle O^{1\dot{2}}(v,u) \rangle = {\rm e}^{\ii\frac{\sqrt{2}v}{R}}A \bar{B}.
\end{equation}

This example highlights a new feature of three-charge microstates: the VEVs of some operators, like $O^{2\dot{1}}$ and $O^{1\dot{2}}$ in our example, are $v$-dependent. This $v$-dependence  originates from the presence of momentum charge (carried in our case by the current $J^+_{-1}$) and from the fact that states dual to geometries are not eigenstates of the momentum operator. Since holography relates the VEVs of operators with the coefficients of the metric expanded around AdS, this implies that three-charge microstate geometries are generically $v$-dependent. 

The geometry dual to $\psi(A, B)$ is given in eqs. (5.2)-(5.3) of \cite{Giusto:2013bda}. At the first non-trivial order in the asymptotic expansion around the AdS boundary, this three-charge solution admits an expansion of the form \eqref{eq:asymptotics}, where the only non-trivial metric functions are
\begin{subequations}
 \begin{align}
Z_4 &\approx  R\,a\,b\, \frac{\cos\theta}{r^3}\, \cos\left(\frac{\sqrt{2}v}{R} - \psi\right)\,,\quad \mathcal{F} \approx -\frac{b^2}{r^2}\,,\\
\beta &\approx \frac{R\,a^2}{\sqrt{2}}\frac{\sin^2\theta\,d\phi-\cos^2\theta\,d\psi}{r^2}\,,\quad \omega \approx \frac{R\,(a^2+b^2)}{\sqrt{2}}\frac{\sin^2\theta\,d\phi+\cos^2\theta\,d\psi}{r^2}\,;
\end{align}
\end{subequations}
the coefficients $a$ and $b$ are taken to be real. The gravity coefficients extracted from this geometry are then 
\begin{align}
\mathcal{A}_{13} + \ii \mathcal{A}_{14}= \frac{R\,a\,b}{2\sqrt{Q_1 Q_5}}\, {\rm e}^{\ii\frac{\sqrt{2}v}{R}}\,,\quad Q_p =  \frac{b^2}{2}\,,\quad a_{3+} = \frac{R\,(a^2+b^2)}{2\sqrt{Q_1Q_5}}\,,\quad  a_{3-} = -\frac{R\,a^2}{2\sqrt{Q_1Q_5}}\,.
\end{align}
Using the dictionary \eqref{eq:dimensionless}, \eqref{eq:dimensionless3c}, along with \eqref{chargequantization}, \eqref{gravity_cft_vev}, and \eqref{eq:cvalues}, we find agreement with the CFT results for the 1-point functions  $\langle O^{2\dot{1}}\rangle$, $\langle O^{1\dot{2}}\rangle$, $\langle J^3\rangle$, $\langle \tilde J^3\rangle$, $\langle \tilde{L}_0-L_0 \rangle$.

\subsection{3 charges and three kinds of strands}\label{sec:3c3s}
The state analyzed in the previous section is a very particular three-charge state: as explained in \cite{Giusto:2013bda}, that state can be generated by acting on the two-charge state with strands $\ket{++}_1$ and $\ket{00}_1$ with the symmetry operator ${\rm e}^{\frac{\pi}{2} (J^+_{-1}- J^-_{+1})}$. We call such states descendants. We consider in this section a simple state which is {\it not} a descendant. This state has also the property that the VEVs of all the dimension one operators are non-trivial and it will allow us to provide a CFT derivation of a numerical coefficient which was  fixed in \cite{Bena:2015bea} by a non-trivial regularity requirement. 

The state we consider has the form (\ref{eq:o3cstate}) with three type of strands: $\ket{++}_1$, $J^+_{-1} \ket{00}_2$, $\ket{00}_1$. We rename the associated coefficients as $A_1^{(++)}\equiv A, B_{2,1}\equiv B_1, B_{1,0}\equiv B_2$ and the respective numbers of strands as $N_1^{(++)}\equiv N-2p_1-p_2$, $N_{2,1}^{(00)}\equiv p_1, N_{1,0}^{(00)}\equiv p_2$, so that the state can be written as
\begin{equation}
\psi(A, B_1, B_2) = \sum_{p_1=0}^{N/2} \sum_{p_2=0}^{N-2p_1} \left( A \ket{++}_1\right)^{N-2p_1-p_2} \left( B_1 J_{-1}^+ \ket{00}_2\right)^{p_1}  \left( B_2 \ket{00}_1\right)^{p_2}\,.
\end{equation}
It is important to keep in mind that the state $J_{-1}^+ \ket{00}_2$ has norm 2
\begin{equation}
_2 \langle 00| J^-_{+1} J^+_{-1}\ket{00}_2 = 2\,,
\end{equation}
as a consequence of the  fractional mode contributions which appear when $J^+_{-1}$ acts on a strand of length 2: $(J^+_{-1})_2 = \psi^{1\dot{1}}_{-1}\psi^{1\dot{2}}_{0}+ \psi^{1\dot{1}}_{0}\psi^{1\dot{2}}_{-1}+ \psi^{1\dot{1}}_{-1/2}\psi^{1\dot{2}}_{-1/2}$. The same mechanism gives rise, for generic $k$ and $m_k$, to the factor $\binom{k}{m_k}$ in (\ref{eq:normNN3c}). We can then borrow the general result \eqref{eq:peak3c} to obtain the average numbers of strands in our state:
\begin{equation}
\bar{p}_1 = |B_1|^2,\qquad \bar{p}_2 = |B_2|^2,\qquad N-2\bar{p}_1 - \bar{p}_2 = |A|^2,
\end{equation}
where the constraint among the coefficients is now
\begin{equation}
\label{constraint_3c3s}
|A|^2 + 2|B_1|^2 + |B_2|^2 = N.
\end{equation}

Since the strands $\ket{++}_1$, $J^+_{-1}\ket{00}_2$ and $\ket{00}_1$ carry spin $(1/2,1/2)$, $(1,0)$ and $(0,0)$, the VEVs of the angular momentum operators are
\begin{equation}
 \label{eq:3c3s_J_VEVs}
\langle J^3 \rangle = \frac{\overline{N}^{(++)}_1}{2} + \overline{N}_{2,1}^{(00)} = \frac{|A|^2}{2} + |B_1|^2,\qquad \langle \tilde{J}^3 \rangle = \frac{\overline{N}^{(++)}_1}{2} = \frac{|A|^2}{2}.
\end{equation}
Momentum is carried only by the $J^+_{-1}\ket{00}_2$ strands, and thus
\begin{equation}
 \label{eq:3c3s_T_VEV}
\langle \tilde{T} \rangle = 0\;,~~~
\langle T \rangle =  \overline{N}_{2,1}^{(00)} = |B_1|^2~~\Rightarrow~~
 n_p = \langle  L_0 - \tilde{L}_0 \rangle = |B_1|^2\,.
\end{equation}

The presence of $\ket{00}_1$ and $\ket{++}_1$ strands signals that some of the $O^{\alpha\dot{\alpha}}$ operators can have a nonzero VEV, while the presence of strands of length 2 with non-zero modes of the current acting on them implies a nonzero $v$-dependent VEV for the twist operators $\Sigma_2^{\alpha\dot{\alpha}}$. On the gravity side, the VEV of $\Sigma_2^{\alpha\dot{\alpha}}$ corresponds to a term of order $r^{-3}$ in the metric function $Z_1$ (see Eq.~(\ref{eq:asymptotics1})); it was shown in \cite{Bena:2015bea} that such a term is needed to ensure regularity of the metric. We will verify that the precise numerical coefficient derived in \cite{Bena:2015bea} matches the CFT prediction. 

Consider first the VEV of $O^{\alpha\dot{\alpha}}$. By angular momentum conservation only $O^{2\dot{2}}$ and $O^{1\dot{1}}$ can acquire a VEV; in particular one can consider the process in which  $O^{2\dot{2}}$ converts a $\ket{++}_1$ into a $\ket{00}_1$ strand:
\begin{align}
O^{2\dot{2}} &\biggl[ \left( \ket{++}_1\right)^{N-2p_1-p_2} \left(  J_{-1}^+ \ket{00}_2\right)^{p_1}  \left(  \ket{00}_1\right)^{p_2} \biggr] = \nonumber\\
&= (p_2+1) \biggl[ \left(  \ket{++}_1\right)^{N-2p_1-p_2-1} \left(  J_{-1}^+ \ket{00}_2\right)^{p_1}  \left(  \ket{00}_1\right)^{p_2+1} \biggr].
\end{align}
This gives rise to the VEV 
\begin{align}
 \label{eq:3c3s_skenderis_VEV_1}
\langle  O^{2\dot{2}} \rangle = \frac{A}{B_2} \bar{p}_2 = A \bar{B}_2\,.
\end{align}
By hermiticity, $O^{1\dot{1}} = \left( O^{2\dot{2}}\right)^\dagger$, one also obtains the VEV
\begin{equation}
 \label{eq:3c3s_skenderis_VEV_2}
\langle O^{1\dot{1}} \rangle=\langle  O^{2\dot{2}} \rangle^*= \bar{A} B_2.
\end{equation}

Consider now $\Sigma_2^{\alpha\dot{\alpha}}$. The twist operator can join two strands of length one into a length two strand of type $J^+_{-1}\ket{00}_2$; by angular momentum conservation, the two starting strands have to be $\ket{++}_1$ and  $\ket{00}_1$ and the operator acting on them $\Sigma_2^{+-}$.  We thus expect the basic correlator  
\begin{equation}
\label{eq:Sigma+-corr}
\phantom{1}_{2} \langle00|\, J^{-}_{+1} \,\Sigma^{+-}_2 \biggl( |++\rangle_{1}\otimes |00\rangle_{1}\biggr)_\mathrm{Symm.}
\end{equation}
to be non-vanishing, where we have denoted by 
\begin{equation}
\label{eq:symmetrized}
 \biggl( |++\rangle_{1}\otimes |00\rangle_{1}\biggr)_\mathrm{Symm.}\equiv |++\rangle_{(r=1)}\otimes |00\rangle_{(r=2)} +  |00\rangle_{(r=1)}\otimes |++\rangle_{(r=2)}
\end{equation}
the product of the two states $\ket{++}_1$ and  $\ket{00}_1$ symmetrized over two copies ($r=1,2$) of the CFT. Note that 
\begin{equation}
\label{eq:norm2}
\left \| \biggl( |++\rangle_{1}\otimes |00\rangle_{1}\biggr)_\text{Symm.} \right \|^2 = 2.
\end{equation}
To compute (\ref{eq:Sigma+-corr}) we need to know the commutator of the doublet of operators $\Sigma_2^{\alpha\dot{\alpha}}$ with the currents $J^i_n$; this has a form analogous to (\ref{eq:commutatorJ}):
\begin{equation}
\left[ \left(J^i_n\right)^{\alpha\beta}, \Sigma_2^{\beta\dot{\alpha}}(v, u)\right] = \frac{1}{2} {\rm e}^{\ii n\frac{\sqrt{2}v}{R}} \left(\sigma^i\right)^{\alpha\beta} \Sigma_2^{\beta\dot{\alpha}}(v,u)\,.
\end{equation}
Thus we find
\begin{align}
\label{eq:Sigma+-corrbis}
\phantom{1}_{2} \langle00|\,J^{-}_{+1} \Sigma^{+-}_2 \biggl( |++\rangle_{1}\otimes |00\rangle_{1}\biggr)_\text{Symm.} \!\!&= {\rm e}^{\ii\frac{\sqrt{2}v}{R}} \phantom{1}_{2} \langle00|\, \Sigma_2^{--} \biggl( |++\rangle_{1}\otimes |00\rangle_{1}\biggr)_\text{Symm.} =  {\rm e}^{\ii\frac{\sqrt{2}v}{R}}\,,
\end{align}
where we have used 
\begin{equation}
\Sigma_2^{--} \biggl( |++\rangle_{1}\otimes |00\rangle_{1}\biggr)_\text{Symm.} = \ket{00}_{2}\,.
\end{equation}
We can now include the combinatorial factors\footnote{The number of ways $\Sigma_2^{+-}$ can act on the strands $\left( \ket{++}_1\right)^{N-2p_1-p_2} \left(  \ket{00}_1\right)^{p_2}$ is $\frac{(N-2p_1-p_2) \,p_2}{2}$; we divide by 2 because we have already taken into account the exchange of a $\ket{++}_1$ and a $\ket{00}_1$ strand in the symmetrized combination (\ref{eq:symmetrized}).} and obtain the action of $\Sigma_2^{+-}$ on the full ensemble of strands:
\begin{align}
\Sigma_2^{+-} &\biggl[ \left( \ket{++}_1\right)^{N-2p_1-p_2} \left(  J_{-1}^+ \ket{00}_2\right)^{p_1}  \left(  \ket{00}_1\right)^{p_2} \biggr] =\nonumber\\
&= \frac{{\rm e}^{\ii\frac{\sqrt{2}v}{R}}}{2} (p_1+1) \biggl[ \left( \ket{++}_1\right)^{N-2p_1-p_2-1} \left(  J_{-1}^+ \ket{00}_2\right)^{p_1+1}  \left(  \ket{00}_1\right)^{p_2-1} \biggr]\,.
\end{align}
Hence we obtain the VEV
\begin{align}
 \label{eq:3c3s_sigma_VEV_1}
\langle \Sigma_2^{+-}(v,u)\rangle = \frac{{\rm e}^{\ii\frac{\sqrt{2}v}{R}}}{2}\, \frac{A B_2}{B_1}\, \bar{p}_1 = \frac{{\rm e}^{\ii\frac{\sqrt{2}v}{R}}}{2} A\, \bar{B}_1 B_2.
\end{align}

Of course one can also consider the opposite process in which $\Sigma_2^{-+}$ acts on $J^+_{-1} |00\rangle_{2}$ to produce singly wound strands $\ket{++}_1 \otimes \ket{00}_1$.  This is captured by
\begin{equation}
\Sigma_2^{-+} J^+_{-1} |00\rangle_{2} = - {\rm e}^{-\ii\frac{\sqrt{2}v}{R}}\,\Sigma_2^{++} \ket{00}_{2}= - \frac{{\rm e}^{-\ii\frac{\sqrt{2}v}{R}}}{2} \,\biggl(  |++\rangle_{1} \otimes |00\rangle_{1} \biggr)_\text{Symm.}   \,,
\end{equation}
where we have used 
\begin{equation}
\Sigma_2^{++} \ket{00}_{2} = \frac{1}{2} \biggl( |++\rangle_{1}\otimes |00\rangle_{1}\biggr)_\text{Symm.}\,.
\end{equation}
Together with (\ref{eq:norm2}) this implies
\begin{equation}
\biggl(\!\!\!\phantom{1}_{1}\langle ++|\otimes\ _{1}\langle 00|\biggr)_\text{Symm.} \Sigma_2^{-+} J^+_{-1} |00\rangle_{2} = - {\rm e}^{-\ii\frac{\sqrt{2}v}{R}}\,,
\end{equation}
which is consistent with (\ref{eq:Sigma+-corrbis}) and the property $\Sigma_2^{-+} = -\left( \Sigma_2^{+-} \right)^\dagger$. Thus
\begin{align}
 \label{eq:3c3s_sigma_VEV_2}
\langle \Sigma_2^{-+}(v,u)\rangle =-\langle \Sigma_2^{+-}(v,u)\rangle^*=- \frac{{\rm e}^{-\ii\frac{\sqrt{2}v}{R}}}{2} \bar{A}\, B_1 \bar{B}_2.
\end{align}

Let us now compare with the dual geometry: it can be read off from Section 5.2 of \cite{Bena:2015bea} taking $k_1=2$, $m_1=1$. We focus in particular on the metric functions $Z_1$, $Z_2$ and $Z_4$, which determine the VEVs of $O^{\alpha\dot\alpha}$ and $\Sigma_2^{\alpha\dot\alpha}$ (the gravity values of the momentum and of the angular momenta are given in (6.11) and (6.15) of  \cite{Bena:2015bea} and are easily seen to agree with the CFT values computed above). At the relevant order in the $1/r$ expansion, the gravity solution is characterized by
\begin{subequations}
\label{R_3c3s}
\begin{align}
Z_1 &= \frac{Q_1}{r^2} + \frac{R^2}{2\,Q_5} \, a\, b_1 b_2\,\cos\left(\frac{\sqrt{2}v}{R}-\psi\right) \frac{\cos\theta}{r^3} +O(r^{-4})\,,\quad Z_2 = \frac{Q_5}{r^2} +O(r^{-4})\,, \\
Z_4 &=R\,a\,b_2\,  \frac{\sin\theta\,\cos\phi}{r^3} +O(r^{-4})\,, \quad\quad \mathcal{F} = - \frac{b_1^2}{4r^2}\,,\\
\beta &= \frac{Ra^2}{\sqrt{2}r^2} \bigl( \sin^2\theta\, d\phi - \cos^2\theta\, d\psi \bigr)\,, \quad \omega = \frac{R \left( a^2 + \frac{b_1^2}{4} \right)}{\sqrt{2}r^2} \bigl( \sin^2\theta\, d\phi + \cos^2\theta\, d\psi \bigr)
\end{align}
\end{subequations}
where the parameters $a,b_1,b_2$ are real. The order $r^{-3}$ term of $Z_1$ is necessary for having a regular metric: its numerical value is determined by the constant $c$ given in Eq.~(5.15) of \cite{Bena:2015bea}, which in our case reads $c=1/2$. Transforming into a coordinate system where (\ref{eq:gauge}) is satisfied, one finds
\begin{subequations}
 \label{eq:gravity3strands}
\begin{align}
&f^1_{13} + \ii f^1_{14} = \frac{R^2}{Q_1 Q_5} \frac{a \,b_1 b_2}{8}\, {\rm e}^{\ii\frac{\sqrt{2}v}{R}}\,,\qquad\quad \mathcal{A}_{11} = \frac{R\,a b_2}{2\sqrt{Q_1 Q_5}}\,,\\
&a_{3+} = \frac{R}{\sqrt{Q_1 Q_5}} \frac{1}{2} \left( a^2 + \frac{b_1^2}{4}\right)\,, \qquad a_{3-} = -\frac{R}{\sqrt{Q_1 Q_5}} \frac{a^2}{2}\,,\quad Q_p= \frac{b_1^2}{8}.
\end{align}
\end{subequations}
Using the holographic dictionary \eqref{gravity_cft_vev}, the values of the constant $c_{O^{(1,1)}}$ and $c_{O^{(0,0)}}$ in (\ref{eq:cvalues}) and (\ref{twist_constant}), and the relations \eqref{eq:dimensionless} and \eqref{eq:dimensionless3c}, which give
\begin{equation}
\label{eq:3c3s_parameters}
A = a \frac{R\sqrt{N}}{\sqrt{Q_1 Q_5}},\qquad B_1 = \frac{b_1}{2\sqrt{2}} \frac{R\sqrt{N}}{\sqrt{Q_1 Q_5}},\qquad B_2 = \frac{b_2}{\sqrt{2}} \frac{R\sqrt{N}}{\sqrt{Q_1 Q_5}},
\end{equation}
one verifies that the gravity result \eqref{eq:gravity3strands} matches, including all numerical factors, with the CFT VEVs \eqref{eq:3c3s_J_VEVs}, \eqref{eq:3c3s_T_VEV}, \eqref{eq:3c3s_skenderis_VEV_1}, \eqref{eq:3c3s_skenderis_VEV_2}, \eqref{eq:3c3s_sigma_VEV_1} and \eqref{eq:3c3s_sigma_VEV_2}.

\section{Entanglement entropy in D1-D5 microstates}
The entanglement entropy of a space domain $A$ in a given microstate represents a useful observable to characterize the microstate itself.  The investigation of this observable, for a domain $A$ composed of a single interval of length $l$ in a two-charge microstate, was initiated in \cite{Giusto:2014aba}, where by following~\cite{Calabrese:2009ez,Calabrese:2010he} it was shown that the EE admits an expansion for small $l$ in terms of the VEVs of operators whose dimensions increase with the order of the $l$-expansion. If only chiral primary operators (and their descendants) are kept in this expansion, the resulting EE coincides with the one evaluated at the gravity point in the CFT moduli space, which, on the other hand, can be holographically computed via the Ryu-Takayanagi formula \cite{Ryu:2006bv} (and its generalizations \cite{Hubeny:2007xt}). Hence the EE provides an alternative handle to compare the VEVs of chiral operators in D1-D5 microstates in the CFT and the gravity pictures. We extend here the results of \cite{Giusto:2014aba} by considering more general two-charge microstates, with non-vanishing VEVs for the twist operators, and also a class of three-charge microstates. 

Before analyzing particular examples, we describe a general approach for the holographic and the CFT derivations of the EE in microstate geometries. 

\subsection{Holographic computation at the first non-trivial order}\label{sec:grav_EE}

The original formalism of Ryu-Takayanagi applies to static asymptotically AdS geometries; as microstate geometries are not static, the appropriate formulation is the covariant one developed in \cite{Hubeny:2007xt}: the EE is given by the area of the co-dimension two surface that extremizes the area functional and is homotopic to the entangling domain $A$, seen as a submanifold of the AdS boundary. Our situation has a further complication, in that microstate geometries are asymptotically AdS$_3\times S^3$ (as the compact space trivially decouples for our class of microstates, we directly work in the 6D Einstein geometry obtained by compactification on $T^4$); moreover generic microstates have a product structure only at the boundary, and there is no invariant way to decouple the AdS$_3$ and the $S^3$ part in the spacetime interior. In \cite{Giusto:2014aba} a recipe was given to write the 6D space as an $S^3$ part fibered over an asymptotically AdS$_3$ space (mathematically, to define an almost product structure); the recipe was based on the introduction of a special set of coordinates, defined, at the first non-trivial order in the expansion around the AdS boundary, by a de Donder gauge condition. This recipe allows to define a 3D asymptotically AdS metric, to which the formulas of \cite{Ryu:2006bv,Hubeny:2007xt} can be directly applied; moreover, reducing the problem from 6D to 3D drastically simplifies the computation of the EE. 

Though the recipe used in \cite{Giusto:2014aba} correctly reproduces the CFT result at the first non-trivial order in the small $l$ expansion, it would be desirable to have an a priori justification for the gauge choice defining the AdS$_3\times S^3$ split. An alternative, geometrically natural, procedure\footnote{We thank R. Emparan, V. Hubeny and J. Simon for drawing our attention to this point.} to holographically compute the EE in spaces that are asymptotically AdS$_3\times S^3$, is to consider, as suggested by \cite{Hubeny:2007xt}, an extramal co-dimension two surface in the full 6D space that reduces at the boundary to $\partial A\times S^3$. We will show here the equivalence between the invariant 6D and the gauge-fixed 3D recipes, at the first non-trivial order in the expansion around the AdS boundary (which coincides with the small $l$ expansion). The extension to higher orders remains an interesting open problem. 

The 6D\footnote{Though we specify to geometries that are asymptotically AdS$_3\times S^3$, as they are the ones relevant for the D1-D5 system, the argument can be readily extended to AdS$\times S$-type of spaces in arbitrary dimension.} Einstein metric can, in full generality, be written in the form
\begin{equation}
\label{eq:S3reduction}
ds^2_6 \equiv  G_{MN}\,dx^M dx^N = g_{\mu\nu}\, dx^\mu dx^\nu + G_{\alpha\beta}(dx^\alpha +A^\alpha_\mu\, dx^\mu)(dx^\beta +A^\beta_\nu \,dx^\nu)\,,
\end{equation}
so that one has
\begin{equation}
  A_\mu^\alpha = G^{\alpha\beta} G_{\mu\beta}~,~~~
  g_{\mu\nu} = G_{\mu\nu} -  G^{\alpha\beta} G_{\mu\alpha} G_{\nu\beta}\;.
\end{equation}
The coordinates are chosen in such a way that $x^\mu$ ($x^\alpha$) are AdS$_3$ ($S^3$) coordinates at the boundary;  the continuation of these coordinates to the interior of the space is, a priori, arbitrary. In \cite{Giusto:2014aba} this arbitrariness was (partly) fixed by requiring that the gauge fields $A^\alpha_\mu$ satisfy the gauge condition
\begin{equation}\label{deDonder}
\nabla^0_\alpha A^\alpha_\mu=0\,,
\end{equation}
with $\nabla^0_\alpha$ the covariant derivative with respect to round metric of $S^3$. We will see that this gauge choice simplifies the covariant EE computation and reduces the problem to the 3D one solved in \cite{Giusto:2014aba}.

In this 6D geometry, consider a co-dimension two submanifold which at the boundary reduces to $S^3$ times a co-dimension two submanifold in AdS$_3$ given by the boundary of the entangling domain $A=[0,l]$. We can parametrize its worldvolume by $x^\alpha$ plus a parameters $\lambda$, so that the parametric representation of the submanifold is $(x^\mu(\lambda,x^\alpha),x^\alpha)$. The metric induced on the submanifold is
\be
ds^2_* =  g_{\mu\nu} dx^\mu_* dx^\nu_* + G_{\alpha\beta}(dx^\alpha +A^\alpha_\mu dx^\mu_*)(dx^\beta +A^\beta_\nu dx^\nu_*)\equiv g^*_{IJ} d\lambda^I d\lambda^J\,,
\ee
with
\be
dx^\mu_* = \dot{x}^\mu d\lambda + \partial_\alpha x^\mu dx^\alpha\,,
\ee
and $\lambda^I \equiv (\lambda, x^\alpha)$. According to the recipe of \cite{Hubeny:2007xt}, this submanifold should extremize the area functional:
\be\label{extremality}
\frac{\partial}{\partial x^\mu} \sqrt{\mathrm{det} g^*} - \frac{\partial}{\partial \lambda} \frac{\partial}{\partial \dot{x}^\mu} \sqrt{\mathrm{det} g^*} - \frac{\partial}{\partial x^\alpha} \frac{\partial}{\partial \partial_\alpha x^\mu} \sqrt{\mathrm{det} g^*}=0\,,
\ee
 where we abbreviate $\dot{x}^\mu \equiv \partial_\lambda x^\mu$. These are complicated partial differential equations for the unknowns $x^\mu(\lambda,x^\alpha)$. However, in the limit of small $l$, the extremal surface probes only a region of spacetime very near the AdS boundary, and, at least at leading order in this asymptotic expansion, the extremality equations can be reduced to simpler ordinary differential equations for functions the functions $X^\mu(\lambda)\equiv \int dx^\alpha \sqrt{\mathrm{det}\,G^0}\,x^\mu(\lambda,x^\alpha)$. To perform this perturbative analysis, we introduce a parameter $\epsilon$ that controls the expansion away from the AdS boundary; the first non-trivial corrections to the metric have the form
\be
\label{eq:asymptoticsEE}
g_{\mu\nu} \equiv g^0_{\mu\nu}+ \epsilon\, \delta g^1_{\mu\nu} + \epsilon^2\, \delta g^2_{\mu\nu}\,,\quad G_{\alpha\beta}\equiv G^0_{\alpha\beta}+ \epsilon\,\delta G^1_{\alpha\beta}+ \epsilon^2\,\delta G^2_{\alpha\beta}\,,\quad A^\alpha_\mu \equiv \epsilon \,\delta A^{\alpha}_\mu\,,
\ee
where $g^0_{\mu\nu}$ is the AdS$_3$ metric, which only depends on $x^\mu$, and $G^0_{\alpha\beta}$ is the $S^3$ metric, which only depends on $x^\alpha$; the correction terms, $\delta g^i_{\mu\nu}$, $\delta G^i_{\alpha\beta}$, $\delta A^\alpha_\mu$, depend both on $x^\mu$ and $x^\alpha$. Correspondingly the functions describing the submanifold can be expanded as
\be\label{xmuexpansion}
x^\mu(\lambda,x^\alpha) = x^\mu_0(\lambda) + \epsilon\, x^\mu_1(\lambda,x^\alpha)+ \epsilon^2 \,x^\mu_2(\lambda, x^\alpha) + O(\epsilon^3)\,,
\ee
where $x^\mu_0(\lambda^i)$ is an extremal surface in AdS$_3$. The expansion (\ref{eq:asymptoticsEE}) descends from the asymptotic expansion (\ref{eq:asymptotics}), where one should think of $f^I_{1i}$, $\mathcal{A}_{1i}$, $a_{\alpha\pm}$ as being proportional to $\epsilon$, while $Q_p$ is proportional to $\epsilon^2$. One can then verify that, for our geometries, the first order corrections to the AdS$_3$ and the $S^3$ metrics vanish:  $\delta g^1_{\mu\nu} =\delta G^1_{\alpha\beta}=0$. Since, as we will see, the gauge fields $A^\alpha_\mu$ only contribute quadratically, this implies that the first non-trivial corrections to the extremal surface $x^\mu(\lambda,x^\alpha)$ and to the EE appear at order $\epsilon^2$. We will limit our analysis to these first non-trivial corrections in this paper. 
 
In Appendix~\ref{Appendix:proofee} we provide the proof of the following facts:

(i) in the gauge (\ref{deDonder}), the first order corrections to the extremal surface vanish: $x^\mu_1(\lambda,x^\alpha)=0$;

(ii) at order $\epsilon^2$ the area of the extremal surface, and hence the EE, only depends on the $S^3$ integral of the extremal surface: $X^\mu(\lambda)\equiv \int dx^\alpha \sqrt{\mathrm{det}\,G^0}\,x^\mu(\lambda,x^\alpha)$;

(iii) the extremality equations for $X^\mu(\lambda)$ are the geodesic equations for a curve in a reduced 3D metric 
\be\label{eq:reduced_3D_metricBIS}
\tilde g_{\mu\nu} \equiv g^0_{\mu\nu}+\epsilon^2 \int \!\!dx^\alpha\, \sqrt{\mathrm{det} G^0} \,\Bigl(\delta g^2_{\mu\nu} + \frac{1}{3} g^0_{\mu\nu} \,G^{\alpha\beta}_0 \delta G^2_{\alpha\beta}\Bigr)\,.
\ee
These are precisely the equations considered in \cite{Giusto:2014aba}. 

\subsection{CFT computation at the first non-trivial order}\label{sec:cft_EE}
The CFT result for the EE for a single interval $A$ of length $l$ at order $\sim l^2$ is
\begin{align}
 \label{eq:EE_cft}
S_A =& \left[ 2N \log\left(\frac{l}{R}\right) - \frac{l^2}{12R^2}\left( -2 \langle T\rangle +\mathcal{N}^{-1}_J \langle J^\alpha\rangle^2 + \mathcal{N}^{-1}_{\tilde{J}} \langle \tilde{J}^\alpha\rangle^2 +\right.\right.\nonumber\\
&\Bigl.\biggl.\quad+ \mathcal{N}^{-1}_{O^{(1,1)}} \langle O^{(1,1)}_{(1)1i}\rangle^2 + \mathcal{N}^{-1}_{O^{(0,0)}} \langle O^{(0,0)}_{(2)i}\rangle^2 \Bigr) + O\left((l/R)^3\right)\biggr],
\end{align}
where the $\mathcal{N}$ coefficients are the normalizations of the two-point functions of the operators
\begin{align}
\langle 0 | J^\alpha(1) J^\beta(0) |0\rangle &= \mathcal{N}_J\, \delta^{\alpha\beta},\qquad\qquad\qquad \langle 0 | \tilde{J}^\alpha(1) \tilde{J}^\beta(0) |0\rangle = \mathcal{N}_{\tilde{J}}\, \delta^{\alpha\beta},\nonumber\\
\langle 0| O^{(1,1)}_{(1)1i} O^{(1,1)}_{(1)1j} |0\rangle &= \mathcal{N}_{O^{(1,1)}} \,\delta_{ij},\qquad\qquad\quad \langle 0| O^{(0,0)}_{(2)i} O^{(0,0)}_{(2)1j} |0\rangle = \mathcal{N}_{O^{(0,0)}}\,\delta_{ij}\,,
\end{align}
with values
\begin{align}
 \label{eq:normalizations}
\mathcal{N}_J = \mathcal{N}_{\tilde{J}} = \mathcal{N}_{O^{(1,1)}} = \frac{n_1 n_5}{2}.
\end{align}
Part of this result was found in \cite{Giusto:2014aba}, the only difference being that here we need to compute the explicit value of $\mathcal{N}_{O^{(0,0)}}$ and we have an extra term coming from the VEV of the stress-energy operator.

The computation of $\mathcal{N}_{O^{(0,0)}}$ is straightforward: it is sufficient to consider a state $(\ket{++}_1)^N$ and compute the VEV of $\Sigma_2^{++} \Sigma_2^{--}$ on it, which by \eqref{eq:Sigma2action} and \eqref{eq:Sigma2daggeraction}  yields
\begin{equation}
\bigl(_1\langle ++|\bigr)^N \Sigma_2^{++} \Sigma_2^{--} \bigl(\ket{++}_1\bigr)^N = \frac{N(N-1)}{2} \simeq \frac{N^2}{2}.
\end{equation}
Writing the operators $\Sigma_2^{\pm\pm}$ in terms of $O^{(0,0)}_{(2)i}$ as in \eqref{eq:sigma2++--} we get an extra factor $1/2$ ($\mathcal{N}_{O^{(0,0)}}$ is defined starting from the real operators) which gives
\begin{equation}
 \label{eq:sigma2_normalization}
\mathcal{N}_{O^{(0,0)}} \simeq \frac{N^2}{4}.
\end{equation}

As explained in \cite{Giusto:2014aba}, all the terms but the one related to $T$ come from contributions of 2-point functions of the CFT primaries: the contributions of the 1-point functions of primaries give zero, and in the case analyzed there no descendants had a nonzero VEV. In the present case, though, $T$ is a descendant of the identity operator, has a nonzero VEV and because of its conformal dimension it gives a contribution of the same order in $l$ as the 2-point functions. This new contribution can be computed exploiting the procedure followed in \cite{Calabrese:2009ez,Calabrese:2010he}. The EE for a single interval $A$ in the dual CFT can be written as
\begin{equation}
 \label{eq:cft_EE_prescription}
S_A = - \frac{\partial}{\partial n} S_n |_{n=1}, \qquad\qquad S_n = \langle s| \mathcal{T}_n(z,\bar{z}) \mathcal{T}_{-n}(w,\bar{w})|s\rangle,
\end{equation}
with
\begin{equation}
\mathcal{T}_n(z,\bar{z}) \mathcal{T}_{-n}(w,\bar{w}) = |z-w|^{-4\Delta_n} \left( 1 + \sum_{K} \sum_{j=1}^n (z-w)^{\Delta_K} (\bar{z}-\bar{w})^{\bar{\Delta}_K} d_K^{(j)} O_K^{(j)} + \cdots \right),
\end{equation}
where we have written only the contribution of single CFT operators on a copy of the CFT (i.e. not tensor products or two or more of them) and $\Delta_n = \bar{\Delta}_n = \frac{c}{24}(n-\frac{1}{n})$ is the conformal dimension of the twist fields $\mathcal{T}_{\pm n}$. We can isolate the contribution given by $T$ multiplying both sides by $T(u)$, taking the VEV and comparing the terms in $\sim (u-w)^{-4}$ as $z\to w$. From the OPE
\begin{equation}
T(z) T(w) \sim \frac{c/2}{(z-w)^4} + \frac{2T(w)}{(z-w)^2} + \frac{\partial T(w)}{z-w}
\end{equation}
we have that the relevant part of the RHS is
\begin{equation}
|z-w|^{-4\Delta_n} (z-w)^2 d_K^{(j)} \frac{c/2}{(u-w)^4}.
\end{equation}
The LHS is given entirely by the nontrivial Schwarzian derivative that we get after we make a conformal transformation to deal with the presence of the twist fields in the 3-point function (see \cite{Calabrese:2004eu}). In the $z\to w$ limit this gives
\begin{equation}
\left(\frac{\Delta_n}{n}\right) \frac{(z-w)^2}{(u-w)^4 |z-w|^{4\Delta_n}},
\end{equation}
so we get
\begin{equation}
d_K^{(j)} = \left( \frac{2}{c} \right) \frac{\Delta_n}{n} = \frac{1}{12} \left( 1- \frac{1}{n^2} \right).
\end{equation}
Notice that within respect to \cite{Calabrese:2004eu}, our $\Delta_n$ is defined after summing over $j$, which gives an extra factor $n$.

The result for the contribution to $S_n$ (where again the sum over $j$ brings just a factor $n$) is
\begin{equation}
S_{n,T} =  |z-w|^{-4\Delta_n} \left(z-w\right)^2 \frac{1}{12} \left( n -\frac{1}{n}\right)\langle s| T(w) |s\rangle,
\end{equation}
which gives a contribution to the EE
\begin{equation}
S_{A,T} = \frac{l^2}{6 R^2} \ \langle s| T(w) |s\rangle.
\end{equation}

\subsection{Entanglement Entropy of three-charge states}
We now want to compare the CFT prediction for the single interval EE derived in Section \ref{sec:cft_EE}, with the holographic computation outlined in Section \ref{sec:grav_EE}. For generic D1-D5-P states, we immediately face the difficulty that we do not know the general expression of the dual geometry. We have however verified, through the examples of Sections \ref{subsection:3c_2st} and \ref{sec:3c3s}, that the 3-charge solutions found in \cite{Bena:2015bea} have an asymptotic expansion of the form \eqref{eq:asymptotics}. We conjecture that this is true for all three-charge states. The knowledge of the expansion \eqref{eq:asymptotics} is enough to compute the EE at to order ${\cal O}(l/R)^2$, and hence compare with the CFT result \eqref{eq:EE_cft}.

Starting with the 6D metric given in \eqref{eq:generalmetric} with the metric coefficients expanded as in  \eqref{eq:asymptotics}, one derives the reduced 3D metric defined in  \eqref{eq:reduced_3D_metricBIS}:
\begin{align}
\tilde{g}_{tt} &= -\frac{r^2}{\sqrt{Q_1 Q_5}} \left[ 1 + 2\delta\mathcal{P} + \frac{1}{r^2}\left( (a_+)^2 + (a_-)^2 -Q_p\right)\right] + O(r^{-2})\,,\\
\tilde{g}_{yy} &= \frac{r^2}{\sqrt{Q_1 Q_5}} \left[ 1 + 2\delta\mathcal{P} - \frac{1}{r^2}\left( (a_+)^2 + (a_-)^2 -Q_p\right)\right]+ O(r^{-2})\,,\\
\tilde{g}_{rr} &= \frac{r^2}{\sqrt{Q_1 Q_5}} \biggl[ 1 + 4\delta\mathcal{P} \biggr]+ O(r^{-2})\,,\\
\tilde{g}_{ty} &= \frac{r^2}{\sqrt{Q_1 Q_5}} \left[ -\frac{1}{r^2}\left(  (a_+)^2 - (a_-)^2 -Q_p\right) \right]+ O(r^{-2})\,,
\end{align}
with
\begin{equation}
\delta \mathcal{P} = -\frac{1}{2} \frac{(f_1^1)^2}{r^2} - \frac{1}{2} \frac{(\mathcal{A}_1^1)^2}{r^2}\,,
\end{equation}
where
\begin{equation}
(a_\pm)^2 \equiv \sum_{\alpha=1}^3 (a_{\alpha\pm})^2\,,\quad (f_1^1)^2 \equiv \sum_{i=1}^4 (f^1_{1i})^2\,,\quad (\mathcal{A}_1)^2 \equiv \sum_{i=1}^4 (\mathcal{A}_{1i})^2\,.
\end{equation}

The gauge fields coming from the reduction on $S^3$ \eqref{eq:S3reduction} are
\begin{equation}
A^\alpha_v = \sqrt{2}\,G^{\alpha\beta}_0 a_{\gamma +} (Y^{\gamma +}_1)_\beta +O(r^{-2})\,,\quad A^\alpha_u = \sqrt{2}\,G^{\alpha\beta}_0 a_{\gamma -} (Y^{\gamma -}_1)_\beta +O(r^{-2})\,,
\end{equation}
with $G^{\alpha\beta}_0$ the inverse of the round $S^3$ metric. They satisfy the gauge condition \eqref{deDonder} because the vector spherical harmonics are divergence-less:
\begin{equation}
\nabla^\alpha (Y^{\gamma \pm}_1)_\alpha =0\,.
\end{equation}
As explained in Section~\ref{sec:grav_EE}, we can thus apply the Ryu-Takayanagi procedure to the reduced 3D metric $\tilde g_{\mu\nu}$ and we obtain the result:
\begin{equation}\label{eq:EEgravity}
S_A  = 2n_1 n_5\!\left[ \log\left( \frac{r_0 l}{\sqrt{Q_1 Q_5}}\right) \!-\! \frac{l^2}{12 Q_1 Q_5}\!\left(-Q_p  +(a_+)^2 + (a_-)^2 +2( \mathcal{A}_1)^2 + 2(f_1^1)^2  \right) + O(l^3)\right] .
\end{equation}
One immediately recognizes a structure similar to \eqref{eq:EE_cft}: the term $(f_1^1)^2$ corresponds to the contribution given by $O^{(0,0)}_{(2)i}\equiv \Sigma_2^{\alpha\dot\alpha}$, the term  $(\mathcal{A}_1)^2$ to $O^{(1,1)}_{(1)1i}\equiv O^{\alpha\dot\alpha}$, the terms $(a_\pm)^2$ to $J^\alpha$ and $\tilde{J}^\alpha$ and the term $Q_p$ to $ \langle L_0 - \tilde{L}_0\rangle$. To verify that also the numerical coefficients match, one uses the relations between the gravity parameters $f^1_{1i}$, $\mathcal{A}_{1i}$, $a_{\alpha\pm}$, $Q_p$ and the CFT VEVs given in \eqref{gravity_cft_vev} with the coefficients
$c_{O^{(0,0)}}$, $c_{O^{(1,1)}}$, $c_J$, $c_{\tilde J}$ specified in \eqref{twist_constant} and \eqref{eq:cvalues}, and the values of the normalization constants $\mathcal{N}$ in \eqref{eq:normalizations} and \eqref{eq:sigma2_normalization}. One can check that these substitutions map precisely the gravity result \eqref{eq:EEgravity} into the CFT formula \eqref{eq:EE_cft}. Part of this match was already performed in \cite{Giusto:2014aba}; what is new here is the momentum contribution proportional to $Q_p\sim \langle  L_0 - \tilde{L}_0 \rangle$ and the verification of the numerical factor in front of the twist field term proportional to $(f_1^1)^2 \sim \langle \Sigma_2 \rangle^2$. Note that this provides an independent non-trivial check of the coefficient $c_{O^{(0,0)}}$, which was fixed in Section \ref{Section:twist} by requiring the CFT-gravity consistency for one particular microstate.\newline
The contribution of $T$ also agrees with the expansion for small $L$ of equation (3.11) of \cite{Caputa:2013lfa} with $r_0^2 = Q_p$.

\section{Discussion and Outlook}
\label{sec:discussion}

The 1-point functions of BPS operators and the single interval EE are useful observables to establish a link between microstates and the dual geometries, and to enlighten the emergence of the spacetime from the CFT. Even if the computations of this paper were limited to chiral primaries of dimension 1 and to the first non-trivial corrections to the EE in the small interval limit, the detailed match between gravity and CFT results provides a quite impressive verification of the map between $1/4$-BPS states and two-charge geometries proposed in \cite{Kanitscheider:2006zf,Kanitscheider:2007wq}, and of its extension to the $1/8$-BPS states of \cite{Bena:2015bea}. In examples like the one worked out in Appendix~\ref{Appendix:twist}, a relatively simple gravity result is matched against a very non-trivial CFT computation,  which uses the correlators of twist operators\footnote{The techniques for handling twist operator insertions in orbifold CFTs have been developed in a long series of papers~\cite{Lunin:2000yv,Carson:2014ena};  the effects of these insertions on the EE have been investigated in \cite{Asplund:2011cq}.} derived in~\cite{Carson:2014ena}. In other examples, like the one of Section~\ref{sec:3c3s},   the presence of a particular term in the geometry follows, in the gravity picture, from a quite involved regularity analysis \cite{Bena:2015bea}, while it is implied quite straightforwardly by the non-vanishing of a twist operator VEV,  in the CFT picture. This last phenomenon is surprising, because the analysis of regularity requires the knowledge of the geometry in the interior of the spacetime, while the CFT picture only involves operators of small dimension (one, in our case), which are associated with deformations of the geometry close to the AdS boundary. This example highlights the power of the CFT in predicting non-trivial features of the dual spacetime. 

Hence, a natural extension of our work consists in extracting from the CFT analysis the necessary information to construct the geometries dual to a larger and more generic family of three-charge states than the one known at present, possibly capturing a finite fraction of the D1-D5-P entropy. In the three-charge microstates of \cite{Bena:2015bea}, the momentum is carried by the current $J^+_{-1}$ acting on strands with spin $(0,0)$; when spectrally flowed to the NS sector, $J^+_{-1}$ becomes $J^+_{0}$ \cite{Mathur:2003hj}. Together with $L_0$, $L_{\pm 1}$,\footnote{Geometries dual to states where momentum is carried by $L_{-1}$ were constructed at linear level in \cite{Mathur:2011gz} and can be extended to nonlinear level using methods similar to \cite{Bena:2015bea}.} the modes $J^\alpha_0$ form the rigid subsector of the CFT chiral algebra, and states where momentum is carried by these rigid generators constitute the so-called ``graviton gas'' contribution to the D1-D5-P elliptic genus \cite{deBoer:1998ip,deBoer:1998us}. The full elliptic genus includes states where momentum is carried by fractional-moded currents acting on strands of winding greater than one: indeed these states dominate the entropy in the limit of large charges. Constructing the geometries dual to such states\footnote{Particular states in this class have already been constructed in \cite{Giusto:2012yz}.} is crucial for the advancement of the so-called ``fuzzball'' program\cite{Mathur:2009hf,Bena:2007kg,Skenderis:2008qn}, which aims at providing a geometric description of black hole microstates in terms supergravity (or more generally string theory) configurations without horizons. For the purpose of this construction, the information provided by the VEVs of BPS operators of dimension larger than one, which determine the higher orders in the asymptotic expansion \eqref{eq:asymptotics}, could be essential. Extending the holographic analysis to higher dimension operators could pose technical hurdles (like the operator mixing phenomenon discussed in \cite{Taylor:2007hs}), but the general methods developed in \cite{Skenderis:2006uy,Skenderis:2006di,Skenderis:2007yb} should allow progress in this direction. 

Having higher dimension operators under control would also be necessary for understanding how a thermal behavior emerges from typical black hole microstates and to quantify the deviations between typical pure states and statistical ensembles \cite{Balasubramanian:2005mg,Balasubramanian:2007qv,Lashkari:2014pna}. The states we consider in this paper are not generic representatives of the ensemble giving rise to the black hole entropy, and indeed the VEVs of simple, low dimension operators, which are non-vanishing in our states, are expected to be suppressed in the large charge limit for typical microstates. But more complex, higher dimension operators can have non-trivial VEVs also in typical states. At least for BPS operators, the free orbifiold CFT picture described in this paper offers a precise tool to characterize and estimate the correlators which can distinguish generic states among themselves and from the maximally mixed state. The holographic dictionary will then allow to determine if and how these differences manifest themselves in the classical geometry. 

Similar questions could be addressed by using the single interval EE as a probe of the microstate geometry. As we have seen, when the length of the interval is small, the EE only probes the region of spacetime close to the boundary, and is only sensitive to operators of small dimension. But as the length increases, the entangling curve reaches deeper in the bulk, possibly exploring the whole spacetime\footnote{The regions of the geometry that are not swept by the entangling curve are called entanglement shadows. The existence of shadows in geometries containing conical defects was pointed out in \cite{Balasubramanian:2014sra}; in the D1-D5 CFT, these geometries are dual to pure states containing multiply wound strands with spin $(\pm 1/2, \pm 1/2)$.}. It has been argued \cite{Asplund:2014coa,Fitzpatrick:2015zha,Perlmutter:2015iya} that in the limit of large central charge, the EE in a typical pure state is dominated by the conformal block of the identity, and hence it reproduces the thermal answer associated with the BTZ black hole \cite{Caputa:2013lfa}. On the other hand we have seen that in our atypical states, the EE receives contributions also from the conformal blocks of non-trivial chiral primaries. It would be interesting to quantify the contribution of non-trivial primaries to the EE in typical states, and evaluate the induced deviations from the thermal behavior.

\vspace{7mm}
 \noindent {\large \textbf{Acknowledgements} }

 \vspace{5mm} 

We would like to thank I. Bena, M. Shigemori and N. Warner for collaboration in a related project, and  R. Emparan, V. Hubeny, V. Jejjala, S. Mathur, J. Simon, K. Skenderis and M. Taylor for useful discussions and correspondence. This research is partially supported by STFC (Grant ST/L000415/1, {\it String theory, gauge theory \& duality}), by the Padova University Project CPDA119349 and by INFN.

\appendix

\section{More details on the orbifold CFT}
\label{Appendix:dCFT}

It is convenient to bosonize the fermions $\psi_\hr$~\eqref{eq:T2eigenvec} that diagonalize the boundary conditions
\begin{equation}
  \label{eq:bosonization}
  \psi^{1 \dot{1}}_{\hr} = \ii {\rm e}^{\ii H_\hr}~,~~~ 
  \psi^{1 \dot{2}}_{\hr} = {\rm e}^{\ii K_\hr}~,
\end{equation}
where $H_\hr$ and $K_\hr$ are compact boson satisfying\footnote{To be precise we would need to include appropriate cocycles so as to ensure that different fermions $\psi^{1 \dot{1}}_{\hr}$ and $\psi^{1 \dot{2}}_{\hr'}$ anticommute also in the bosonic language. We will skip this detail.} the OPE
\begin{equation}
  \label{eq:bosOPE}
  H_\hr(z) H_{\hr'}(w) \sim K_\hr(z) K_{\hr'}(w) \sim -\delta_{\hr \hr'} \ln(z-w)\;,~~
  H_\hr(z) K_{\hr'}(w) \sim 0\;.
\end{equation}
The fermionic part of the twist fields introduced in Section~\ref{sect:dCFT} can be written in terms of the free fields $H$ and $K$ in a standard way; for instance
\begin{equation}
\label{Sibos}
  \Sigma^{-\frac{k-1}{2},-\frac{k-1}{2}}_k = \otimes_{\hr=0}^{{k-1}} \left( \sigma_\hr^{X} \; {\rm e}^{-\ii\frac{\hr}{k} H_\hr}\, {\rm e}^{-\ii\frac{\hr}{k} K_\hr}\right)  \left(\tilde{\sigma}_\hr^{X} \; {\rm e}^{-\ii\frac{\hr}{k} \tilde{H}_\hr}\, {\rm e}^{-\ii\frac{\hr}{k} \tilde{K}_\hr}\right)~,
\end{equation}
where $\sigma_\hr^{X}$ is the bosonic twist field acting on the bosonic coordinates $X^{A\dot{A}}_\hr$. The conformal weight of $\Sigma_k$ is 
$$
\sum_{\hr=0}^{k-1} \left[\frac{\hr}{k} \left(1 - \frac{\hr}{k} \right) +  \frac{\hr^2}{k^2} \right] = \frac{k-1}{2}~
$$
and similarly for the anti-holomorphic part: the first term in the sum is the $\sigma_\hr^{X}$ contribution and the second one is the contribution from the fermionic sector. In order to calculate explicitly the action of the twist fields on a R ground state such as $\otimes_r \ket{++}_{(r)} \equiv \ket{++}^k$, we need to rewrite the state in terms of the bosons introduced here. We have simply
\begin{equation}
  \label{rrkHK}
  \ket{++}^k = \lim_{z\to 0} \otimes_{\hr=0}^{k-1} \left({\rm e}^{\frac{\ii}{2} ({H}_\hr(z) + K_\hr(z))} {\rm e}^{\frac{\ii}{2} (\tilde{H}_\hr(\bar{z})+\tilde{K}_\hr(\bar{z}))} \right) \ket{0}~,
\end{equation}
where $\ket{0}$ is the $SL(2,C)$ invariant vacuum. Eq.~\eqref{rrkHK} can be checked simply by verifying that all $\psi_{(r)}$'s have the R monodromies and that the state defined has the expected spin and conformal dimension. Then it is straightforward to calculate the r.h.s. of~\eqref{sSi} by using free fields
\begin{align}
 \label{eq:SgsR}
\Sigma^{-\frac{k-1}{2},-\frac{k-1}{2}}_k (z,\bar{z}) \ket{++}^k \sim&  |z|^{-(k-1)} \otimes_{\hr=0}^{{k-1}} \left[ \sigma^X_\rho {\rm e}^{\ii\left(-\frac{\hr}{k}+\frac{1}{2}\right) H_\hr(0)} {\rm e}^{\ii\left(-\frac{\hr}{k}+\frac{1}{2}\right) K_\hr(0)}\right]\times\nonumber\\
&\times \left[ \tilde{\sigma}^X_\rho {\rm e}^{\ii\left(-\frac{\hr}{k}+\frac{1}{2}\right) \tilde{H}_\hr(0)} {\rm e}^{\ii\left(-\frac{\hr}{k}+\frac{1}{2}\right) \tilde{K}_\hr(0)}\right] |0\rangle~.
\end{align}
Notice that the state produced is again part of the R sector. This can be seen by relating the periodicity of bulk fermions along the $S^1$ parametrized by $y$ to the periodicity of the CFT fermions. For instance we can consider the bulk gravitino field that couples to the CFT super-current, $G(z)$ and the two objects are either both periodic or both antiperiodic. If we consider $k$ copies of the CFT, the total supercurrent is the sum of the supercurrents of the individual copies is $\sum_r G_{(r)} = G_{\hr=0}$, where, as in~\eqref{eq:T2eigenvec}, $G_\hr$ are the components in the twisted sectors that diagonlize the boundary conditions we have
\begin{equation}
  \label{eq:Geigenvec}
   G_\hr (z) = \sum_{r=1}^k {\rm e}^{2\pi \ii \frac{r \hr}{k}}G_{r}(z)~,~~~
  \mbox{with}~~\hr=0,1,\ldots,k-1~.
\end{equation}
The fermionic fields $G_{\hr=0}(z)$ have the same mode expansions as $\psi_{\hr=0}(z)$, so when it goes around the NS state we have
\begin{equation}\label{GNSm}
G_{\hr=0}({\rm e}^{2\pi \ii} z) = G_{\hr=0}(z)\,,
\end{equation}
while when the current goes around a R state we have
\begin{equation}\label{GRm}
G_{\hr=0}({\rm e}^{2\pi \ii} z) = -G_{\hr=0} (z)\,.
\end{equation}
The state obtained by the action of $\Sigma_k$ in~\eqref{eq:SgsR} has the monodromies in~\eqref{GRm} and so is part of the R sector.

\section{General gravity results for D1-D5 geometries}
\label{Appendix:D1D5geometries}
We will now give some general results for the objects $Z_1, Z_2$ and $Z_4$ for 2-charge geometries up to order $\sim 1/r^3$. First we define
\begin{equation}
h_1(v') \equiv g_1(v') + \ii g_2(v'), \qquad h_2(v')\equiv g_3(v') + \ii g_4(v').
\end{equation}

We have
\begin{align}
Z_1 &\equiv 1+ \frac{Q_5}{L} \int_0^L dv \frac{|\dot{h}_1|^2 + |\dot{h}_2|^2 + |\dot{g}_5|^2}{|x_i-g_i|^2},\quad Z_2 \equiv 1 + \frac{Q_5}{L} \int_0^L dv \frac{1}{|x_i-g_i|^2},\\
\mathcal{A} &\equiv -\frac{Q_5}{L}\int_0^L dv \frac{\dot{g}_5}{|x_i-g_i|^2},\quad\qquad\qquad\quad\quad A \equiv -\frac{Q_5}{L} \int_0^L dv \frac{\dot{g}_j dx^j}{|x_i-g_i|^2},
\end{align}
where the denominator can also be rewritten as
\begin{equation}
|x_i-g_i|^2\equiv \sum_{i=1}^4 \left( x_i-g_i\right)^2 = |(x_1+\ii x_2)-h_1|^2 + |(x_3+\ii x_4)-h_2|^2.
\end{equation}

The result for $Z_1$ at order $\sim 1/r^3$ is
\begin{align}
Z_1 &\simeq \frac{4\pi^2 Q_5}{L^2} \sum_{k\neq 0}\left\lbrace |a^{(1)}_k|^2 + |a^{(2)}_k|^2 + \frac{1}{4} |a^{(00)}_{|k|}|^2\right\rbrace+\nonumber\\
&\quad+\frac{4\pi^2 Q_5}{L^2} \frac{1}{r^3} \sum_{k,l,n\neq 0} \frac{1}{n} \left\lbrace \left(\frac{x_1-\ii x_2}{r}\right) \left( a^{(1)}_k \bar{a}^{(1)}_l a^{(1)}_n \delta_{k-l+n} + a^{(2)}_k \bar{a}^{(2)}_l a^{(1)}_n \delta_{k-l+n} \right)+\right.\nonumber\\
&\quad\left. + \left(\frac{x_3-\ii x_4}{r}\right)\left( a^{(1)}_k \bar{a}^{(1)}_l a^{(2)}_n \delta_{k-l+n}  + a^{(2)}_k \bar{a}^{(2)}_l a^{(2)}_n \delta_{k-l+n} \right) + [\mathrm{c.c.}]  \right\rbrace+\nonumber\\
&\quad +\frac{\pi^2 Q_5}{L^2} \frac{1}{r^3} \sum_{k,l=0}^{+\infty} \sum_{n\neq 0} \frac{1}{n}\left\lbrace \left(\frac{x_1-\ii x_2}{r}\right) \left( a^{(00)}_k a^{(00)}_l a^{(1)}_n \delta_{k+l+n} + 2a^{(00)}_k \bar{a}^{(00)}_l a^{(1)}_n \delta_{k-l+n} +\right.\right.\nonumber\\
&\left.\left. \quad+  \bar{a}^{(00)}_k \bar{a}^{(00)}_l a^{(1)}_n \delta_{-k-l+n}\right)+\right.\nonumber\\
&\left.\quad+ \left(\frac{x_3-\ii x_4}{r}\right) \left( a^{(00)}_k a^{(00)}_l a^{(2)}_n \delta_{k+l+n} + 2a^{(00)}_k \bar{a}^{(00)}_l a^{(2)}_n \delta_{k-l+n} +\right.\right.\nonumber\\
&\biggl.\left. \quad +  \bar{a}^{(00)}_k \bar{a}^{(00)}_l a^{(2)}_n \delta_{-k-l+n}\right)+ [\mathrm{c.c.}]\biggr\rbrace,
\end{align}
where $\delta_{m}\equiv \delta_{m,0}$ and where we put $a^{(00)}_{k<0} = 0$.

The result for $Z_2$ does not contain terms of order $\sim 1/r^3$, thus
\begin{equation}
Z_2 = 1 + \frac{Q_5}{r^2} + O\left(\frac{1}{r^4}\right),
\end{equation}
while for $Z_4$ we have
\begin{align}
Z_4 =& \frac{\pi Q_5}{L} \frac{1}{r^3} \sum_{k=1}^{+\infty} \frac{1}{k} \left\lbrace \left(\frac{x_1-\ii x_2}{r}\right) \left( a_k^{(00)} a_{-k}^{(1)} + \bar{a}_k^{(00)} a_{k}^{(1)}\right) +\right.\nonumber\\
&\left.+\left(\frac{x_3-\ii x_4}{r}\right) \left( a_k^{(00)} a_{-k}^{(2)} + \bar{a}_k^{(00)} a_{k}^{(2)}\right) + [\mathrm{c.c.}] \right\rbrace.
\end{align}

The 1-form $A=A_i dx^i$ can be written at order $\sim 1/r^3$ as
\begin{equation}
A_i = -2Q_5 f_{ij} \frac{x_j}{r^4},\qquad f_{ij}\equiv \frac{1}{L}\int_0^L dv\ \dot{g}_i g_j = - f_{ji}.
\end{equation}
We can switch to complex coordinates
\begin{align}
z_1 \equiv x_1+\ii x_2, \qquad \bar{z}_1\equiv x_1-\ii x_2,\\
z_2 \equiv x_3+\ii x_4, \qquad \bar{z}_2\equiv x_3-\ii x_4,
\end{align}
and define indices $z^a, z^b,\ldots$ such that $z^a=(z_1, \bar{z}_1, z_2, \bar{z}_2)$ and so on to get
\begin{equation}
A_{z^a} = -2Q_5 f_{z^a z^b} \frac{dz^b}{r^4}.
\end{equation}
We have
\begin{align}
f_{z_1 \bar{z}_1} &= \frac{2\pi i}{L} \sum_{n\neq 0} a_n^{(1)} \frac{\bar{a}_n^{(1)}}{n},\qquad f_{z_1 z_2} = -\left(f_{\bar{z}_1 \bar{z}_2}\right)^* = -\frac{2\pi i}{L} \sum_{n\neq 0} a_n^{(1)} \frac{a_{-n}^{(2)}}{n},\\
f_{z_2 \bar{z}_2} &= \frac{2\pi i}{L} \sum_{n\neq 0} a_n^{(2)} \frac{\bar{a}_n^{(2)}}{n},\qquad f_{z_1 \bar{z}_2} = -\left( f_{\bar{z}_1, z_2} \right)^* = \frac{2\pi i}{L} \sum_{n\neq 0} a_n^{(1)} \frac{\bar{a}_n^{(2)}}{n}.
\end{align}

The components of the 1-form $B$ at order $\sim 1/r^3$ are obtained in the coordinates $x_i$ as
\begin{equation}
B_i = -Q_5\ \epsilon_{ijkl} f_{kl} \frac{x_j}{r^4}.
\end{equation}

\section{General D1-D5 state with twist field VEV}
\label{Appendix:twist}
In general the twist field $\Sigma_2^{\alpha\dot\alpha}$ can join two strands of length $k_1$ and $k_2$ into a strand or length $k_1+k_2$ (or vice versa split the $k_1+k_2$-long strand into $k_1$ and $k_2$ pieces). Thus a state with three different strands of lengths $k_1, k_2$ and $k_3 = k_1 + k_2$ will have a non-vanishing VEV for $\Sigma_2^{\alpha\dot\alpha}$. For simplicity we take the spin state of all the strands to be $(++)$, so our building blocks are $\ket{++}_{k_i}$, with $i=1,2,3$. In Section~\ref{Section:twist} we have considered the particular case with $k_1=k_2=1$, $k_3=2$. The interest of the more general case relies on the fact that the action of the twist field on strands of length greater than one is quite subtle, and it produces a non-trivial numerical factor which was computed by CFT methods in \cite{Carson:2014ena} (see Eq.~(5.25) there). We will show that holography provides a non-trivial check for this coefficient. 

The state we consider has the form (\ref{eq:gravityCFTmap}) where the only non-trivial coefficients are $A^{(++)}_{k_1}\equiv A_1$, $A^{(++)}_{k_2} \equiv A_2$, $A^{(++)}_{k_3} \equiv A_3$; for brevity, we also rename $N^{(++)}_{k_1} \equiv p_1$, $N^{(++)}_{k_2} \equiv p_2$, $N^{(++)}_{k_3} \equiv p_3$; these numbers are subject to the constraint  $k_1 p_1 + k_2 p_2 + k_3 p_3 = N$. The state is then
\begin{equation}\label{eq:state123}
\psi(A_1, A_2, A_3) \equiv \sum_{p_1=0}^{N/k_1} \sum_{p_2=0}^{\frac{N-k_1p_1}{k_2}}  \left( A_1 \ket{++}_{k_1}\right)^{p_1} \left( A_2 \ket{++}_{k_2}\right)^{p_2} \left( A_3 \ket{++}_{k_3}\right)^{\frac{N-k_1p_1-k_2 p_2}{k_3}}.
\end{equation}
Its norm is
\begin{equation}
|\psi(A_1, A_2, A_3)|^2 = \sum_{p_1=0}^{N/k_1} \sum_{p_2=0}^{\frac{N-k_1p_1}{k_2}} A_1^{p_1} A_2^{p_2}  A_3^{\frac{N-k_1p_1-k_2 p_2}{k_3}}\,\mathcal{N}(p_1,p_2)\,,
\end{equation}
with
\begin{equation}\label{eq:normgeneral}
\mathcal{N}(p_1,p_2) =\frac{N!}{p_1! \,p_2!\, (\frac{N-k_1p_1-k_2p_2}{k_3})!\,k_1^{p_1}\, k_2^{p_2}\, k_3^{\frac{N-k_1p_1-k_2p_2}{k_3}}}\,.
\end{equation}
 According to the general result \eqref{eq:peak} the sum in (\ref{eq:state123}) is peaked around the average values
\begin{equation}
\bar{p}_i = \frac{|A_i|^2}{k_i} \qquad (i=1,2,3)\,.
\end{equation}

We can now consider the action of the twist field on the state $\psi(A,B)$. For angular momentum conservation, only the operator  $\Sigma_2^{--}$ can glue two strands and only  $\Sigma_2^{++}$ can split one strand. The novelty with respect to the state with $k_1=k_2=1$ is that when $\Sigma_2^{(--)}$ glues two strands of windings $k_1, k_2>1$, the final state is multiplied by the factor
\begin{equation}
c_{k_1 k_2} = \frac{k_1+ k_2}{2k_1 k_2}\,.
\end{equation}
Note that $c_{1,1}=1$, and thus this effect was invisible in the computation of Section~\ref{Section:twist}. This factor was derived via a non-trivial CFT computation in \cite{Carson:2014ena}; we will import their result here, and show that it is necessary for consistency with the holographic computation of the VEV. One has moreover to include the usual combinatorial factors which arise when one has multiple strands of the same type, so the total action of the twist field is 
\begin{align}\label{eq:Sigma2actiongenaral}
&\Sigma_2^{--} \biggl[\left( \ket{++}_{k_1}\right)^{p_1} \left( \ket{++}_{k_3}\right)^{p_2} \left( \ket{++}_{k_3}\right)^{p_3}\biggr] =\nonumber\\
&\qquad= c_{k_1k_2}\,(p_3 + 1)\,k_3\,\biggl[ \left( \ket{++}_{k_1}\right)^{p_1-1} \left( \ket{++}_{k_2}\right)^{p_2-1} \left( \ket{++}_{k_3}\right)^{p_3+1}\biggr] \,.
\end{align}
The combinatorics is explained as follows: there are $p_1$ ($p_2$) ways to pick one strand of length $k_1$ ($k_2$); moreover on a strand of length $k_1$ ($k_2$), the gluing action of $\Sigma_2^{--}$ can be applied at $k_1$ ($k_2$) positions within the strand. Thus the number of terms appearing on the l.h.s. of (\ref{eq:Sigma2actiongenaral}) is
\begin{equation}
\label{eq:identitygeneral}
p_1 \,p_2\,k_1\,k_2 \,\mathcal{N}(p_1,p_2) = (p_3 + 1)\,k_3\,\mathcal{N}(p_1-1,p_2-1)\,,
\end{equation}
where we have used (\ref{eq:normgeneral}). Since this equals the number of terms present on the r.h.s. of (\ref{eq:Sigma2actiongenaral}) (up to the factor $c_{k_1,k_2}$), this justifies the combinatorial factors in that equation. 

The calculation for the VEV of $\Sigma_2^{--}$ on $\psi(A_1, A_2, A_3)$ now proceeds along similar lines as in Eq.~(\ref{eq:VEVSigma2}), and one obtains
\begin{align}
\label{eq:VEVSigma2general}
\langle \Sigma_2^{--} \rangle&\equiv  | \psi(A_1, A_2,A_3)|^{-2}\,\langle \psi(A_1,A_2,A_3)| \Sigma_2^{--}\ket{\psi(A_1,A_2,A_3)} \nonumber\\
&=c_{k_1 k_2} \frac{A_1 A_2}{A_3} \bar{p}_3 = \frac{k_1+k_2}{2\,k_1 k_2} A_1 A_2 \bar{A}_3\,.
\end{align}

Analogous arguments determine the action of $\Sigma_2^{++}$, when it splits a strand of winding $k_1+k_2$ into pieces of length $k_1$ and $k_2$:
\begin{align}
&\Sigma_2^{++} \biggl[\left( \ket{++}_{k_1}\right)^{p_1} \left( \ket{++}_{k_3}\right)^{p_2} \left( \ket{++}_{k_3}\right)^{p_3}\biggr] =\nonumber\\
&\qquad= c_{k_1k_2}\,(p_1 + 1)\,k_1\,(p_2+1)\,k_2\,\biggl[ \left( \ket{++}_{k_1}\right)^{p_1+1} \left( \ket{++}_{k_2}\right)^{p_2+1} \left( \ket{++}_{k_3}\right)^{p_3-1}\biggr]\,.
\end{align}
One can again check that, thanks to the identity (\ref{eq:identitygeneral}), the action of $\Sigma_2^{++}$ is consistent with hermitian conjugation and thus
\begin{align}
\langle \Sigma_2^{++}\rangle = \langle \Sigma_2^{--}\rangle^*= \frac{k_1 + k_2 }{2\,k_1 k_2} \bar{A}_1 \bar{A}_2 A_3\,.
\end{align}

The VEVs of the angular momentum operators are determined by the average numbers of strands, and are given by
\begin{align}
\label{eq:angulargeneral}
\langle  J^3 \rangle = \langle  \tilde{J}^3\rangle  = \frac{1}{2} \left( \bar{p}_1 + \bar{p}_2 + \bar{p}_3 \right) = \frac{1}{2} \left( \frac{|A_1|^2}{k_1} + \frac{|A_2|^2}{k_2} + \frac{|A_3|^2}{k_1+k_2} \right)\,.
\end{align}

On the gravity side, the dual geometry is associated with the profile with modes $a_{k_1}^{(++)}\equiv a_1$, $a_{k_2}^{(++)}\equiv a_3$ and $a_{k_3}^{(++)}\equiv a_3$, related with the CFT parameters as 
\begin{align}
\label{eq:dictionarygeneral}
a_i = \frac{A_i}{R} \sqrt{\frac{Q_1 Q_5}{N}}  \qquad (i=1,2,3)\,.
\end{align}

The gravity coefficients determining the VEVs are  
\begin{align}
f^1_{11} - \ii f^1_{12}  &= \frac{R^2}{Q_1 Q_5}\, \frac{k_1+k_2}{2\,k_1 k_2} \,a_1 \, a_2\,{\bar a_3}\,,\quad \mathcal{A}_{1i} = 0\,,\\
a_{3+} = -a_{3-} &= \frac{R}{2\,\sqrt{Q_1 Q_5}}\, \left( \frac{|a_1|^2}{k_1} + \frac{|a_2|^2}{k_2} + \frac{|a_3|^2}{k_3}\right)\,.
\end{align}

The angular momenta derived from $a_{3+}$, $a_{3-}$ are easily seen to match with the CFT values (\ref{eq:angulargeneral}). Using the coefficient $c_{O^{(00)}}$ given in (\ref{twist_constant}),  the gravity prediction for the VEV of $\Sigma_2^{--}$ is
\begin{equation}
\langle \Sigma_2^{--} \rangle_{\mathrm{Grav.}} = c_{O^{(0,0)}}\, (f^1_{11} - \ii f^1_{12})= \frac{N^{3/2}\,R^3}{(Q_1 Q_5)^{3/2}}\, \frac{k_1+k_2}{2\,k_1 k_2} a_1 \,a_2\,{\bar a_3} \,,
\end{equation}
which matches with the CFT prediction (\ref{eq:VEVSigma2general}) in view of (\ref{eq:dictionarygeneral}).

\section{The proof of the statements in Section~\ref{sec:grav_EE}}
\label{Appendix:proofee}

Below we sketch the proof for the statements at the end of Section~\ref{sec:grav_EE}.

(i) Consider the extremality equation (\ref{extremality}) at first order in $\epsilon$. Since $\partial_\alpha x^\mu$ starts at order $\epsilon$, and in (\ref{extremality}) there appears the first derivative of $\sqrt{\mathrm{det} g^*}$ with respect to $\partial_\alpha x^\mu$, it is enough to compute $\sqrt{\mathrm{det} g^*}$ at second order in $\partial_\alpha x^\mu$. This can be done by doing an expansion around $\partial_\alpha x^\mu=0$, where the induced metric $g^*$ greatly simplifies. Indeed when $\partial_\alpha x^\mu=0$ one has
\be
g^*_{\lambda\lambda}= \hat g_{\mu\nu}\, \dot{x}^\mu \dot{x}^\nu\,,\quad g^*_{\lambda\alpha} = G_{\alpha\beta}\,A^\beta_\mu \,\dot{x}^\mu\,,\quad g^*_{\alpha\beta}=G_{\alpha\beta}\,,\quad (\partial_\alpha x^\mu=0)
\ee
with 
\be
\hat g_{\mu\nu}\equiv g_{\mu\nu}+G_{\alpha\beta}A^\alpha_\mu A^\beta_{\nu}\,.
\ee
Then the inverse of the induced metric is
\be
g_*^{\lambda\lambda} = g^{\lambda\lambda}\,,\quad g_*^{\lambda\alpha} =  - g^{\lambda\lambda} \,A^\alpha_\mu \,\dot{x}^\mu\,,\quad g_*^{\alpha\beta} = G^{\alpha\beta}+g^{\lambda\lambda}\, A^\alpha_\mu A^\beta_\nu\, \dot{x}^\mu \dot{x}^\nu\,,\quad (\partial_\alpha x^\mu=0)
\ee
where $g^{\lambda\lambda}$ is the inverse of
\be
g_{\lambda\lambda}\equiv g_{\mu\nu}\, \dot{x}^\mu \dot{x}^\nu\,.
\ee
Using this observation, one can compute the expansion of $\sqrt{\mathrm{det} g^*}$ up to the first order in $\partial_\alpha x^\mu$:
\be\label{orderzeropartialalpha}
\sqrt{\mathrm{det} g^*}\Bigl |_{\partial_\alpha x^\mu=0}= \sqrt{g_{\lambda\lambda}\, \mathrm{det} G}\,,
\ee

\be\label{firstderivative}
 \frac{\partial \sqrt{\mathrm{det} g^*}}{\partial \partial_\alpha x^\mu} \Bigl |_{\partial_\alpha x^\mu=0}=\sqrt{g_{\lambda\lambda}\,\mathrm{det}G}\, (A^\alpha_\mu  - g^{\lambda\lambda} A^\alpha_\sigma  g_{\mu\rho} \dot{x}^\rho \dot{x}^\sigma) \,.
\ee
 
When evaluating the first two terms in the extremality equation (\ref{extremality}) at first order in $\epsilon$, one only needs (\ref{orderzeropartialalpha}); moreover, due to the absence of first order corrections to $g_{\mu\nu}$ and $G_{\alpha\beta}$,  one can approximate
 \be
 \sqrt{\mathrm{det} g^*} =\sqrt{g^0_{\mu\nu} \dot{x}^\mu \dot{x}^\nu}\, \sqrt{\mathrm{det} G^0} + O(\epsilon^2)\,.
 \ee
 Substituting the expansion (\ref{xmuexpansion}) for $x^\mu(\lambda,x^\alpha)$ in the above equation, one immediately concludes that, at first order in $\epsilon$, the first two terms in (\ref{extremality}) give a linear and homogeneous equation for $x^\mu_1$. Consider now the last term in (\ref{extremality}): the only contribution that is not homogeneous in $x^\mu_1$ comes from (\ref{firstderivative}). At our order of approximation such a term is
 \be
 - \frac{\partial}{\partial x^\alpha} \frac{\partial}{\partial \partial_\alpha x^\mu} \sqrt{\mathrm{det} g^*} = -\epsilon \sqrt{g_{\lambda\lambda}^0\,\mathrm{det}G^0}\,g^{\lambda\lambda}_0(\nabla^0_\alpha \delta A^\alpha_\mu\,  g^0_{\rho\sigma} - \nabla^0_\alpha \delta A^\alpha_\sigma\, g^0_{\mu\rho})\, \dot{x}_0^\rho\dot{x}_0^\sigma + O(\epsilon^2)\,,
\ee
where
\be
g^0_{\lambda\lambda}\equiv g^0_{\mu\nu}\, \dot{x}_0^\mu \dot{x}_0^\nu
\ee
does not depend on $x^\alpha$. This term vanishes thanks to the de Donder gauge condition (\ref{deDonder}). We thus conclude that the equation for $x^\mu_1$ is linear and homogeneous and hence it admits the solution  $x^\mu_1=0$.

(ii) Consider now the contributions of order $\epsilon^2$ to the area of the extremal surface
\be
A = \int d\lambda dx^\alpha \sqrt{\mathrm{det} g^*} \,,
\ee
which gives the EE. We notice that to compute $\sqrt{\mathrm{det} g^*}$ up to order $\epsilon^2$ one can set $\partial_\alpha x^\mu=0$: indeed, having shown that $x^\mu_1=0$, we know that $\partial_\alpha x^\mu$ starts at order $\epsilon^2$; moreover (\ref{firstderivative}) implies that the first derivative of $\sqrt{\mathrm{det} g^*}$ with respect to $\partial_\alpha x^\mu$ is at least of order $\epsilon$; thus the contributions from $\partial_\alpha x^\mu$ to $\sqrt{\mathrm{det} g^*}$ are at least of order $\epsilon^3$. For the computation of $A$ we can then use the simplified expression (\ref{orderzeropartialalpha}), and obtain
\be\label{areasecondorder}
A= \int d\lambda dx^\alpha  \sqrt{g_{\lambda\lambda}\, \mathrm{det} G} + O(\epsilon^3)= A_0+\epsilon^2\!\!\int d\lambda \sqrt{g_{\lambda\lambda}^0} \,g^{\lambda\lambda}_0\, g^0_{\mu\nu}\, \dot{x}_0^\mu\, \dot{X}_2^\nu + \ldots + O(\epsilon^3)\,,
\ee
where $A_0$ is the order zero term, $X_2^\mu$ is the $S^3$ integral of $x_2^\mu$ 
\be
X_2^\mu\equiv \int \!\!dx^\alpha\, \sqrt{\mathrm{det} G^0} \,x_2^\mu\,,  
\ee 
and the dots in (\ref{areasecondorder}) are terms of order $\epsilon^2$ that do not depend on $x_2^\mu$ (but are proportional to $\delta g^2_{\mu\nu}$ and $\delta G^2_{\alpha\beta}$). We conclude that to compute $A$ at second order we do not need to know $x_2^\mu(\lambda,x^\alpha)$ but only its integral $X^\mu_2(\lambda)$.

(iii) We now want to derive a differential equation for $X^\mu_2(\lambda)$, or equivalently for $X^\mu(\lambda)$. Since the extremality equation (\ref{extremality}) at order $\epsilon^2$ is of course linear in $x_2^\mu$, we can derive an equation for its $S^3$-integral by integrating (\ref{extremality}) on $S^3$; the last term in (\ref{extremality}), being a total derivative with respect to $x^\alpha$, drops out of the integral; so we get the equation 
\be
 \label{eq:minimal}
\int dx^\alpha \Bigl[\frac{\partial}{\partial x^\mu} \sqrt{\mathrm{det} g^*} - \frac{\partial}{\partial \lambda} \frac{\partial}{\partial \partial_i x^\mu} \sqrt{\mathrm{det} g^*} \Bigr]=0\,,
\ee
where we can use the approximation (\ref{orderzeropartialalpha}) for $\sqrt{\mathrm{det} g^*}$. 

We thus see that the problem reduces to that of finding an extremal surface in the ``reduced 3D'' metric $g^E_{\mu\nu}\equiv g_{\mu\nu} \,(\mathrm{det} G)$. Note that $g^E_{\mu\nu}$ would be the Einstein metric in 3D if it were independent of $x^\alpha$. In this extremality problem the variables $x^\alpha$ appear as external parameters, i.e. the equation does not contain derivatives with respect to $x^\alpha$. At the end of the computation one should integrate over $x^\alpha$. Alternatively one can perform the integral over $x^\alpha$ before solving the equations and define an $x^\alpha$-independent 3D metric
\be
 \label{eq:reduced_3D_metric}
\tilde g_{\mu\nu} \equiv g^0_{\mu\nu}+\epsilon^2 \int \!\!dx^\alpha\, \sqrt{\mathrm{det} G^0} \,\Bigl(\delta g^2_{\mu\nu} + \frac{1}{3} g^0_{\mu\nu} \,G^{\alpha\beta}_0 \delta G^2_{\alpha\beta}\Bigr)\,.
\ee
(Note: we are assuming the normalization $\int \!\!dx^\alpha\, \sqrt{\mathrm{det} G^0} =1$).
The equations that determine $X^\mu(\lambda)\equiv \int \!\!dx^\alpha\, \sqrt{\mathrm{det} G^0} \,x^\mu(\lambda,x^\alpha)$ are the geodesic equations for a curve in the metric $\tilde g_{\mu\nu}$.

\providecommand{\href}[2]{#2}\begingroup\raggedright\endgroup


\begin{thebibliography}{10}

\bibitem{Maldacena:1997re}
J.~M. Maldacena, ``{The large N limit of superconformal field theories and
  supergravity},'' {\em Adv. Theor. Math. Phys.} {\bfseries 2} (1998) 231--252,
\href{http://arxiv.org/abs/hep-th/9711200}{{\ttfamily arXiv:hep-th/9711200}}.

\bibitem{Strominger:1996sh}
A.~Strominger and C.~Vafa, ``{Microscopic origin of the Bekenstein-Hawking
  entropy},'' \href{http://dx.doi.org/10.1016/0370-2693(96)00345-0}{{\em
  Phys.Lett.} {\bfseries B379} (1996) 99--104},
  \href{http://arxiv.org/abs/hep-th/9601029}{{\ttfamily arXiv:hep-th/9601029
  [hep-th]}}.

\bibitem{Callan:1996dv}
C.~G. Callan and J.~M. Maldacena, ``{D-brane approach to black hole quantum
  mechanics},'' \href{http://dx.doi.org/10.1016/0550-3213(96)00225-8}{{\em
  Nucl.Phys.} {\bfseries B472} (1996) 591--610},
  \href{http://arxiv.org/abs/hep-th/9602043}{{\ttfamily arXiv:hep-th/9602043
  [hep-th]}}.

\bibitem{Baggio:2012rr}
M.~Baggio, J.~de~Boer, and K.~Papadodimas, ``{A non-renormalization theorem for
  chiral primary 3-point functions},''
  \href{http://dx.doi.org/10.1007/JHEP07(2012)137}{{\em JHEP} {\bfseries 1207}
  (2012) 137},
\href{http://arxiv.org/abs/1203.1036}{{\ttfamily arXiv:1203.1036 [hep-th]}}.

\bibitem{Kanitscheider:2006zf}
I.~Kanitscheider, K.~Skenderis, and M.~Taylor, ``{Holographic anatomy of
  fuzzballs},'' \href{http://dx.doi.org/10.1088/1126-6708/2007/04/023}{{\em
  JHEP} {\bfseries 0704} (2007) 023},
\href{http://arxiv.org/abs/hep-th/0611171}{{\ttfamily arXiv:hep-th/0611171
  [hep-th]}}.

\bibitem{Kanitscheider:2007wq}
I.~Kanitscheider, K.~Skenderis, and M.~Taylor, ``{Fuzzballs with internal
  excitations},'' {\em JHEP} {\bfseries 06} (2007) 056,
\href{http://arxiv.org/abs/0704.0690}{{\ttfamily arXiv:0704.0690 [hep-th]}}.

\bibitem{Taylor:2007hs}
M.~Taylor, ``{Matching of correlators in AdS(3) / CFT(2)},''
  \href{http://dx.doi.org/10.1088/1126-6708/2008/06/010}{{\em JHEP} {\bfseries
  0806} (2008) 010},
\href{http://arxiv.org/abs/0709.1838}{{\ttfamily arXiv:0709.1838 [hep-th]}}.

\bibitem{Giusto:2004id}
S.~Giusto, S.~D. Mathur, and A.~Saxena, ``{Dual geometries for a set of
  3-charge microstates},''
  \href{http://dx.doi.org/10.1016/j.nuclphysb.2004.09.001}{{\em Nucl. Phys.}
  {\bfseries B701} (2004) 357--379},
\href{http://arxiv.org/abs/hep-th/0405017}{{\ttfamily arXiv:hep-th/0405017}}.

\bibitem{Giusto:2004ip}
S.~Giusto, S.~D. Mathur, and A.~Saxena, ``{3-charge geometries and their CFT
  duals},'' \href{http://dx.doi.org/10.1016/j.nuclphysb.2005.01.009}{{\em Nucl.
  Phys.} {\bfseries B710} (2005) 425--463},
\href{http://arxiv.org/abs/hep-th/0406103}{{\ttfamily arXiv:hep-th/0406103}}.

\bibitem{Ford:2006yb}
J.~Ford, S.~Giusto, and A.~Saxena, ``{A class of BPS time-dependent 3-charge
  microstates from spectral flow},''
  \href{http://dx.doi.org/10.1016/j.nuclphysb.2007.09.008}{{\em Nucl. Phys.}
  {\bfseries B790} (2008) 258--280},
\href{http://arxiv.org/abs/hep-th/0612227}{{\ttfamily arXiv:hep-th/0612227}}.

\bibitem{Bena:2005va}
I.~Bena and N.~P. Warner, ``{Bubbling supertubes and foaming black holes},''
  \href{http://dx.doi.org/10.1103/PhysRevD.74.066001}{{\em Phys.Rev.}
  {\bfseries D74} (2006) 066001},
  \href{http://arxiv.org/abs/hep-th/0505166}{{\ttfamily arXiv:hep-th/0505166
  [hep-th]}}.

\bibitem{Berglund:2005vb}
P.~Berglund, E.~G. Gimon, and T.~S. Levi, ``{Supergravity microstates for BPS
  black holes and black rings},''
  \href{http://dx.doi.org/10.1088/1126-6708/2006/06/007}{{\em JHEP} {\bfseries
  0606} (2006) 007}, \href{http://arxiv.org/abs/hep-th/0505167}{{\ttfamily
  arXiv:hep-th/0505167 [hep-th]}}.

\bibitem{Bena:2006is}
I.~Bena, C.-W. Wang, and N.~P. Warner, ``{The Foaming three-charge black
  hole},'' \href{http://dx.doi.org/10.1103/PhysRevD.75.124026}{{\em Phys.Rev.}
  {\bfseries D75} (2007) 124026},
  \href{http://arxiv.org/abs/hep-th/0604110}{{\ttfamily arXiv:hep-th/0604110
  [hep-th]}}.

\bibitem{Bena:2006kb}
I.~Bena, C.-W. Wang, and N.~P. Warner, ``{Mergers and Typical Black Hole
  Microstates},'' \href{http://dx.doi.org/10.1088/1126-6708/2006/11/042}{{\em
  JHEP} {\bfseries 11} (2006) 042},
\href{http://arxiv.org/abs/hep-th/0608217}{{\ttfamily arXiv:hep-th/0608217}}.

\bibitem{Bena:2015bea}
I.~Bena, S.~Giusto, R.~Russo, M.~Shigemori, and N.~P. Warner, ``{Habemus
  Superstratum! A constructive proof of the existence of superstrata},''
  \href{http://dx.doi.org/10.1007/JHEP05(2015)110}{{\em JHEP} {\bfseries 1505}
  (2015) 110},
\href{http://arxiv.org/abs/1503.01463}{{\ttfamily arXiv:1503.01463 [hep-th]}}.

\bibitem{Calabrese:2004eu}
P.~Calabrese and J.~L. Cardy, ``{Entanglement entropy and quantum field
  theory},'' \href{http://dx.doi.org/10.1088/1742-5468/2004/06/P06002}{{\em
  J.Stat.Mech.} {\bfseries 0406} (2004) P06002},
\href{http://arxiv.org/abs/hep-th/0405152}{{\ttfamily arXiv:hep-th/0405152
  [hep-th]}}.

\bibitem{Ryu:2006bv}
S.~Ryu and T.~Takayanagi, ``{Holographic derivation of entanglement entropy
  from AdS/CFT},'' \href{http://dx.doi.org/10.1103/PhysRevLett.96.181602}{{\em
  Phys.Rev.Lett.} {\bfseries 96} (2006) 181602},
\href{http://arxiv.org/abs/hep-th/0603001}{{\ttfamily arXiv:hep-th/0603001
  [hep-th]}}.

\bibitem{Ryu:2006ef}
S.~Ryu and T.~Takayanagi, ``{Aspects of Holographic Entanglement Entropy},''
  \href{http://dx.doi.org/10.1088/1126-6708/2006/08/045}{{\em JHEP} {\bfseries
  0608} (2006) 045},
\href{http://arxiv.org/abs/hep-th/0605073}{{\ttfamily arXiv:hep-th/0605073
  [hep-th]}}.

\bibitem{Giusto:2014aba}
S.~Giusto and R.~Russo, ``{Entanglement Entropy and D1-D5 geometries},''
  \href{http://dx.doi.org/10.1103/PhysRevD.90.066004}{{\em Phys.Rev.}
  {\bfseries D90} no.~6, (2014) 066004},
\href{http://arxiv.org/abs/1405.6185}{{\ttfamily arXiv:1405.6185 [hep-th]}}.

\bibitem{Hubeny:2007xt}
V.~E. Hubeny, M.~Rangamani, and T.~Takayanagi, ``{A Covariant holographic
  entanglement entropy proposal},''
  \href{http://dx.doi.org/10.1088/1126-6708/2007/07/062}{{\em JHEP} {\bfseries
  0707} (2007) 062},
\href{http://arxiv.org/abs/0705.0016}{{\ttfamily arXiv:0705.0016 [hep-th]}}.

\bibitem{Carson:2014ena}
Z.~Carson, S.~Hampton, S.~D. Mathur, and D.~Turton, ``{Effect of the
  deformation operator in the D1D5 CFT},''
  \href{http://dx.doi.org/10.1007/JHEP01(2015)071}{{\em JHEP} {\bfseries 1501}
  (2015) 071},
\href{http://arxiv.org/abs/1410.4543}{{\ttfamily arXiv:1410.4543 [hep-th]}}.

\bibitem{Jevicki:1998bm}
A.~Jevicki, M.~Mihailescu, and S.~Ramgoolam, ``{Gravity from CFT on S**N(X):
  Symmetries and interactions},''
  \href{http://dx.doi.org/10.1016/S0550-3213(00)00147-4}{{\em Nucl.Phys.}
  {\bfseries B577} (2000) 47--72},
\href{http://arxiv.org/abs/hep-th/9907144}{{\ttfamily arXiv:hep-th/9907144
  [hep-th]}}.

\bibitem{Lunin:2001jy}
O.~Lunin and S.~D. Mathur, ``{AdS/CFT duality and the black hole information
  paradox},'' \href{http://dx.doi.org/10.1016/S0550-3213(01)00620-4}{{\em Nucl.
  Phys.} {\bfseries B623} (2002) 342--394},
\href{http://arxiv.org/abs/hep-th/0109154}{{\ttfamily arXiv:hep-th/0109154}}.

\bibitem{Lunin:2002bj}
O.~Lunin, S.~D. Mathur, and A.~Saxena, ``{What is the gravity dual of a chiral
  primary?},'' \href{http://dx.doi.org/10.1016/S0550-3213(03)00081-6}{{\em
  Nucl. Phys.} {\bfseries B655} (2003) 185--217},
\href{http://arxiv.org/abs/hep-th/0211292}{{\ttfamily arXiv:hep-th/0211292}}.

\bibitem{Lunin:2002iz}
O.~Lunin, J.~M. Maldacena, and L.~Maoz, ``{Gravity solutions for the D1-D5
  system with angular momentum},''
\href{http://arxiv.org/abs/hep-th/0212210}{{\ttfamily arXiv:hep-th/0212210}}.

\bibitem{Skenderis:2006ah}
K.~Skenderis and M.~Taylor, ``{Fuzzball solutions and D1-D5 microstates},''
  \href{http://dx.doi.org/10.1103/PhysRevLett.98.071601}{{\em Phys.Rev.Lett.}
  {\bfseries 98} (2007) 071601},
\href{http://arxiv.org/abs/hep-th/0609154}{{\ttfamily arXiv:hep-th/0609154
  [hep-th]}}.

\bibitem{Bena:2011dd}
I.~Bena, S.~Giusto, M.~Shigemori, and N.~P. Warner, ``{Supersymmetric Solutions
  in Six Dimensions: A Linear Structure},''
  \href{http://dx.doi.org/10.1007/JHEP03(2012)084}{{\em JHEP} {\bfseries 1203}
  (2012) 084},
\href{http://arxiv.org/abs/1110.2781}{{\ttfamily arXiv:1110.2781 [hep-th]}}.

\bibitem{Giusto:2013rxa}
S.~Giusto, L.~Martucci, M.~Petrini, and R.~Russo, ``{6D microstate geometries
  from 10D structures},''
  \href{http://dx.doi.org/10.1016/j.nuclphysb.2013.08.018}{{\em Nucl.Phys.}
  {\bfseries B876} (2013) 509--555},
\href{http://arxiv.org/abs/1306.1745}{{\ttfamily arXiv:1306.1745 [hep-th]}}.

\bibitem{Giusto:2013bda}
S.~Giusto and R.~Russo, ``{Superdescendants of the D1D5 CFT and their dual
  3-charge geometries},'' \href{http://dx.doi.org/10.1007/JHEP03(2014)007}{{\em
  JHEP} {\bfseries 1403} (2014) 007},
\href{http://arxiv.org/abs/1311.5536}{{\ttfamily arXiv:1311.5536 [hep-th]}}.

\bibitem{Calabrese:2009ez}
P.~Calabrese, J.~Cardy, and E.~Tonni, ``{Entanglement entropy of two disjoint
  intervals in conformal field theory},''
  \href{http://dx.doi.org/10.1088/1742-5468/2009/11/P11001}{{\em J.Stat.Mech.}
  {\bfseries 0911} (2009) P11001},
\href{http://arxiv.org/abs/0905.2069}{{\ttfamily arXiv:0905.2069 [hep-th]}}.

\bibitem{Calabrese:2010he}
P.~Calabrese, J.~Cardy, and E.~Tonni, ``{Entanglement entropy of two disjoint
  intervals in conformal field theory II},''
  \href{http://dx.doi.org/10.1088/1742-5468/2011/01/P01021}{{\em J.Stat.Mech.}
  {\bfseries 1101} (2011) P01021},
\href{http://arxiv.org/abs/1011.5482}{{\ttfamily arXiv:1011.5482 [hep-th]}}.

\bibitem{Lunin:2000yv}
O.~Lunin and S.~D. Mathur, ``{Correlation functions for M(N)/S(N) orbifolds},''
  \href{http://dx.doi.org/10.1007/s002200100431}{{\em Commun. Math. Phys.}
  {\bfseries 219} (2001) 399--442},
\href{http://arxiv.org/abs/hep-th/0006196}{{\ttfamily arXiv:hep-th/0006196}}.

O.~Lunin and S.~D. Mathur, ``{Three-point functions for M(N)/S(N) orbifolds
  with N = 4 supersymmetry},''
  \href{http://dx.doi.org/10.1007/s002200200638}{{\em Commun. Math. Phys.}
  {\bfseries 227} (2002) 385--419},
\href{http://arxiv.org/abs/hep-th/0103169}{{\ttfamily arXiv:hep-th/0103169}}.

S.~G. Avery, B.~D. Chowdhury, and S.~D. Mathur, ``{Deforming the D1D5 CFT away
  from the orbifold point},''
  \href{http://dx.doi.org/10.1007/JHEP06(2010)031}{{\em JHEP} {\bfseries 06}
  (2010) 031},
\href{http://arxiv.org/abs/1002.3132}{{\ttfamily arXiv:1002.3132 [hep-th]}}.

S.~G. Avery, B.~D. Chowdhury, and S.~D. Mathur, ``{Excitations in the deformed
  D1D5 CFT},'' \href{http://dx.doi.org/10.1007/JHEP06(2010)032}{{\em JHEP}
  {\bfseries 06} (2010) 032},
\href{http://arxiv.org/abs/1003.2746}{{\ttfamily arXiv:1003.2746 [hep-th]}}.

B.~A. Burrington, A.~W. Peet, and I.~G. Zadeh, ``{Twist-nontwist correlators in
  $M^N/S_N$ orbifold CFTs},''
  \href{http://dx.doi.org/10.1103/PhysRevD.87.106008}{{\em Phys.Rev.}
  {\bfseries D87} no.~10, (2013) 106008},
\href{http://arxiv.org/abs/1211.6689}{{\ttfamily arXiv:1211.6689 [hep-th]}}.

B.~A. Burrington, A.~W. Peet, and I.~G. Zadeh, ``{Operator mixing for string
  states in the D1-D5 CFT near the orbifold point},''
  \href{http://dx.doi.org/10.1103/PhysRevD.87.106001}{{\em Phys.Rev.}
  {\bfseries D87} no.~10, (2013) 106001},
\href{http://arxiv.org/abs/1211.6699}{{\ttfamily arXiv:1211.6699 [hep-th]}}.

Z.~Carson, S.~Hampton, S.~D. Mathur, and D.~Turton, ``{Effect of the twist
  operator in the D1D5 CFT},''
  \href{http://dx.doi.org/10.1007/JHEP08(2014)064}{{\em JHEP} {\bfseries 1408}
  (2014) 064},
\href{http://arxiv.org/abs/1405.0259}{{\ttfamily arXiv:1405.0259 [hep-th]}}.

Z.~Carson, S.~D. Mathur, and D.~Turton, ``{Bogoliubov coefficients for the
  twist operator in the D1D5 CFT},''
  \href{http://dx.doi.org/10.1016/j.nuclphysb.2014.10.018}{{\em Nucl.Phys.}
  {\bfseries B889} (2014) 443--485},
\href{http://arxiv.org/abs/1406.6977}{{\ttfamily arXiv:1406.6977 [hep-th]}}.

B.~A. Burrington, S.~D. Mathur, A.~W. Peet, and I.~G. Zadeh, ``{Analyzing the
  squeezed state generated by a twist deformation},''
  \href{http://dx.doi.org/10.1103/PhysRevD.91.124072}{{\em Phys.Rev.}
  {\bfseries D91} no.~12, (2015) 124072},
\href{http://arxiv.org/abs/1410.5790}{{\ttfamily arXiv:1410.5790 [hep-th]}}.

\bibitem{Asplund:2011cq}
C.~T. Asplund and S.~G. Avery, ``{Evolution of Entanglement Entropy in the
  D1-D5 Brane System},''
  \href{http://dx.doi.org/10.1103/PhysRevD.84.124053}{{\em Phys.Rev.}
  {\bfseries D84} (2011) 124053},
\href{http://arxiv.org/abs/1108.2510}{{\ttfamily arXiv:1108.2510 [hep-th]}}.

\bibitem{Mathur:2003hj}
S.~D. Mathur, A.~Saxena, and Y.~K. Srivastava, ``{Constructing `hair' for the
  three charge hole},''
  \href{http://dx.doi.org/10.1016/j.nuclphysb.2003.12.022}{{\em Nucl.Phys.}
  {\bfseries B680} (2004) 415--449},
\href{http://arxiv.org/abs/hep-th/0311092}{{\ttfamily arXiv:hep-th/0311092
  [hep-th]}}.

\bibitem{deBoer:1998ip}
J.~de~Boer, ``{Six-dimensional supergravity on S**3 x AdS(3) and 2-D conformal
  field theory},'' \href{http://dx.doi.org/10.1016/S0550-3213(99)00160-1}{{\em
  Nucl.Phys.} {\bfseries B548} (1999) 139--166},
\href{http://arxiv.org/abs/hep-th/9806104}{{\ttfamily arXiv:hep-th/9806104
  [hep-th]}}.

\bibitem{deBoer:1998us}
J.~de~Boer, ``{Large N elliptic genus and AdS / CFT correspondence},''
  \href{http://dx.doi.org/10.1088/1126-6708/1999/05/017}{{\em JHEP} {\bfseries
  9905} (1999) 017},
\href{http://arxiv.org/abs/hep-th/9812240}{{\ttfamily arXiv:hep-th/9812240
  [hep-th]}}.

\bibitem{Giusto:2012yz}
S.~Giusto, O.~Lunin, S.~D. Mathur, and D.~Turton, ``{D1-D5-P microstates at the
  cap},'' \href{http://dx.doi.org/10.1007/JHEP02(2013)050}{{\em JHEP}
  {\bfseries 1302} (2013) 050},
\href{http://arxiv.org/abs/1211.0306}{{\ttfamily arXiv:1211.0306 [hep-th]}}.

\bibitem{Mathur:2009hf}
S.~D. Mathur, ``{The information paradox: A pedagogical introduction},''
  \href{http://dx.doi.org/10.1088/0264-9381/26/22/224001}{{\em Class. Quant.
  Grav.} {\bfseries 26} (2009) 224001},
\href{http://arxiv.org/abs/0909.1038}{{\ttfamily arXiv:0909.1038 [hep-th]}}.

\bibitem{Bena:2007kg}
I.~Bena and N.~P. Warner, ``{Black holes, black rings and their microstates},''
  \href{http://dx.doi.org/10.1007/978-3-540-79523-0}{{\em Lect. Notes Phys.}
  {\bfseries 755} (2008) 1--92},
\href{http://arxiv.org/abs/hep-th/0701216}{{\ttfamily arXiv:hep-th/0701216}}.

\bibitem{Skenderis:2008qn}
K.~Skenderis and M.~Taylor, ``{The fuzzball proposal for black holes},''
  \href{http://dx.doi.org/10.1016/j.physrep.2008.08.001}{{\em Phys. Rept.}
  {\bfseries 467} (2008) 117--171},
\href{http://arxiv.org/abs/0804.0552}{{\ttfamily arXiv:0804.0552 [hep-th]}}.

\bibitem{Skenderis:2006uy}
K.~Skenderis and M.~Taylor, ``{Kaluza-Klein holography},''
  \href{http://dx.doi.org/10.1088/1126-6708/2006/05/057}{{\em JHEP} {\bfseries
  0605} (2006) 057},
\href{http://arxiv.org/abs/hep-th/0603016}{{\ttfamily arXiv:hep-th/0603016
  [hep-th]}}.

\bibitem{Skenderis:2006di}
K.~Skenderis and M.~Taylor, ``{Holographic Coulomb branch vevs},''
  \href{http://dx.doi.org/10.1088/1126-6708/2006/08/001}{{\em JHEP} {\bfseries
  0608} (2006) 001},
\href{http://arxiv.org/abs/hep-th/0604169}{{\ttfamily arXiv:hep-th/0604169
  [hep-th]}}.

\bibitem{Skenderis:2007yb}
K.~Skenderis and M.~Taylor, ``{Anatomy of bubbling solutions},''
  \href{http://dx.doi.org/10.1088/1126-6708/2007/09/019}{{\em JHEP} {\bfseries
  0709} (2007) 019}, \href{http://arxiv.org/abs/0706.0216}{{\ttfamily
  arXiv:0706.0216 [hep-th]}}.

\bibitem{Balasubramanian:2005mg}
V.~Balasubramanian, J.~de~Boer, V.~Jejjala, and J.~Simon, ``{The library of
  Babel: On the origin of gravitational thermodynamics},'' {\em JHEP}
  {\bfseries 12} (2005) 006,
\href{http://arxiv.org/abs/hep-th/0508023}{{\ttfamily arXiv:hep-th/0508023}}.

\bibitem{Balasubramanian:2007qv}
V.~Balasubramanian, B.~Czech, V.~E. Hubeny, K.~Larjo, M.~Rangamani, {\em
  et~al.}, ``{Typicality versus thermality: An Analytic distinction},''
  \href{http://dx.doi.org/10.1007/s10714-008-0606-8}{{\em Gen.Rel.Grav.}
  {\bfseries 40} (2008) 1863--1890},
\href{http://arxiv.org/abs/hep-th/0701122}{{\ttfamily arXiv:hep-th/0701122
  [hep-th]}}.

\bibitem{Lashkari:2014pna}
N.~Lashkari and J.~Simón, ``{From state distinguishability to effective bulk
  locality},'' \href{http://dx.doi.org/10.1007/JHEP06(2014)038}{{\em JHEP}
  {\bfseries 1406} (2014) 038},
\href{http://arxiv.org/abs/1402.4829}{{\ttfamily arXiv:1402.4829 [hep-th]}}.

\bibitem{Balasubramanian:2014sra}
V.~Balasubramanian, B.~D. Chowdhury, B.~Czech, and J.~de~Boer, ``{Entwinement
  and the emergence of spacetime},''
  \href{http://dx.doi.org/10.1007/JHEP01(2015)048}{{\em JHEP} {\bfseries 1501}
  (2015) 048},
\href{http://arxiv.org/abs/1406.5859}{{\ttfamily arXiv:1406.5859 [hep-th]}}.

\bibitem{Asplund:2014coa}
C.~T. Asplund, A.~Bernamonti, F.~Galli, and T.~Hartman, ``{Holographic
  Entanglement Entropy from 2d CFT: Heavy States and Local Quenches},''
  \href{http://dx.doi.org/10.1007/JHEP02(2015)171}{{\em JHEP} {\bfseries 1502}
  (2015) 171},
\href{http://arxiv.org/abs/1410.1392}{{\ttfamily arXiv:1410.1392 [hep-th]}}.

\bibitem{Fitzpatrick:2015zha}
A.~L. Fitzpatrick, J.~Kaplan, and M.~T. Walters, ``{Virasoro Conformal Blocks
  and Thermality from Classical Background Fields},''
\href{http://arxiv.org/abs/1501.05315}{{\ttfamily arXiv:1501.05315 [hep-th]}}.

\bibitem{Perlmutter:2015iya}
E.~Perlmutter, ``{Virasoro conformal blocks in closed form},''
\href{http://arxiv.org/abs/1502.07742}{{\ttfamily arXiv:1502.07742 [hep-th]}}.

\bibitem{Caputa:2013lfa}
P.~Caputa, V.~Jejjala, and H.~Soltanpanahi, ``{Entanglement Entropy of Extremal
  BTZ},'' \href{http://dx.doi.org/10.1103/PhysRevD.89.046006}{{\em Phys.Rev.}
  {\bfseries D89} (2014) 046006},
\href{http://arxiv.org/abs/1309.7852}{{\ttfamily arXiv:1309.7852 [hep-th]}}.

\bibitem{Mathur:2011gz}
  S.~D.~Mathur and D.~Turton,
  JHEP {\bf 1205} (2012) 014
  [arXiv:1112.6413 [hep-th]].

\end{thebibliography}
\end{document}